\documentclass[aps,10pt,pra,superscriptaddress,showpacs,floatfix,twocolumn,longbibliography]{revtex4-1}

\usepackage{graphicx}
\usepackage{epstopdf}
\usepackage{amsmath}
\usepackage{comment}
\usepackage{amssymb}
\usepackage{hyperref}
\usepackage{bm}
\usepackage{subfigure}

\hypersetup{
    colorlinks=true,       
    linkcolor=cyan,          
    citecolor=magenta,        
    filecolor=magenta,      
    urlcolor=cyan,           
    runcolor=cyan
}
\usepackage[pdftex]{color}

\newcommand{\beq}{\begin{equation}}
\newcommand{\eeq}{\end{equation}}
\newcommand{\beqnn}{\begin{equation*}}
\newcommand{\eeqnn}{\end{equation*}}
\newcommand{\bea}{\begin{eqnarray}}
\newcommand{\eea}{\end{eqnarray}}
\newcommand{\beann}{\begin{eqnarray*}}
\newcommand{\eeann}{\end{eqnarray*}}
\newcommand{\bes} {\begin{subequations}}
\newcommand{\ees} {\end{subequations}}

\newcommand{\ket}[1]{ | #1\rangle}
\newcommand{\bra}[1]{\langle #1 | }

\newcommand{\ident}{\openone}

\newcommand{\veps}{\varepsilon}

\newcommand{\ignore}[1]{}

\newcommand{\rmd}{{\text d}}

\begin{document}

\title{Analog Nature of Quantum Adiabatic Unstructured Search}

\author{Mikhail Slutskii}
\affiliation{National Research University Higher School of Economics, 101000 Moscow, Russia}

\author{Tameem Albash}
\affiliation{Department of Electrical and Computer Engineering,  University of New Mexico, Albuquerque, New Mexico 87131, USA}
\affiliation{Department of Physics and Astronomy and Center for Quantum Information and Control, CQuIC, University of New Mexico, Albuquerque, New Mexico 87131, USA}

\author{Lev Barash}
\affiliation{National Research University Higher School of Economics, 101000 Moscow, Russia}
\affiliation{Landau Institute for Theoretical Physics, 142432 Chernogolovka, Russia}

\author{Itay Hen}
\email{itayhen@isi.edu}
\affiliation{Department of Physics and Astronomy, and Center for Quantum Information Science \& Technology, University of Southern California, Los Angeles, California 90089, USA}
\affiliation{Information Sciences Institute, University of Southern California, Marina del Rey, California 90292, USA}
\begin{abstract}
The quantum adiabatic unstructured search algorithm is one of only a handful of quantum adiabatic optimization algorithms to exhibit provable speedups over their classical counterparts. With no fault tolerance theorems to guarantee the resilience of such algorithms against errors, understanding the impact of imperfections on their performance is of both scientific and practical significance. We study the robustness of the algorithm against various types of imperfections: limited control over the interpolating schedule, Hamiltonian misspecification, and interactions with a thermal environment. We find  that the unstructured search algorithm's quadratic speedup is generally not robust to the presence of any one of the above non-idealities, and in some cases we find that it imposes unrealistic conditions on how the strength of these noise sources must scale to maintain the quadratic speedup. 
\end{abstract}

\maketitle
%


\section{Introduction}

Quantum adiabatic optimization (QAO)~\cite{Finnila1994343,Brooke30041999,kadowaki:98,farhi:01,santoro:02,howPowerful,Reichardt:2004:QAO:1007352.1007428} is a paradigm of computing in which a slowly time-evolving Hamiltonian that uses continuously decreasing quantum fluctuations is employed 
in order to prepare the ground state of a target Hamiltonian in an analog, rather than digital, manner~\cite{young:08,young:10,hen:11,hen:12,farhi:12,PhysRevA.74.060304,PhysRevA.73.022329}. As such, it is expected by many to be a simpler way of carrying out quantum-assisted calculations experimentally~\cite{vandersypen:01,gaitan:12,Babbush:2014,speedup,DWave,Hen:2015rt,q-sig,q-sig2,DWave,LL1,LL2,LL3,PhysRevA.96.042322,Perdomo-Ortiz:2011fh}. 

To date, there is only a handful of quantum adiabatic optimization algorithms whose runtime is provably superior to their classical counterparts~\cite{roland:02,hen:14,hen:14b,henPeriodFinding,sommaGlued}\footnote{We exclude here algorithms derived via the polynomial equivalence theorem between the circuit and adiabatic models of quantum computing~\cite{aharonov_adiabatic_2007} since the `final' Hamiltonians are not necessarily diagonal in the computational basis.}. First and foremost among these is the quantum adiabatic unstructured search (QAUS) algorithm --- an oracular algorithm for identifying a marked state in an unstructured list. Originally devised by Roland and Cerf~\cite{roland:02} (but see also Refs.~\cite{analogAnalogue,howPowerful} for earlier variants), the algorithm consists of encoding the search space in a `problem Hamiltonian,' $H_p$, that is constant across the entire search space except for one `marked' configuration $\ket{m}=\ket{m_1 m_2 \ldots m_n}$ whose cost is lower than the rest. Here, $m_i \in \{0,1\}$ are the bits of the $n$-bit marked configuration (the number of elements in the search space is thus $N=2^  n$ where $n$ is the number of elements). Similar to its gate-based counterpart, Grover's unstructured search algorithm~\cite{grover:97}, the runtime of the Roland and Cerf algorithm scales as $O(\sqrt{N})$, 
which is to be contrasted with the linear scaling with $N$ of the number of queries required classically for finding the marked item.

While the asymptotic scaling of the runtime of quantum adiabatic algorithms such as QAUS give an accounting of the `time resources' used by the algorithm, one should be careful about not accounting for other resources, particularly precision, due to the analog nature of the algorithm~\cite{Aaronson,VERGIS198691,analogComputing}. Failing to do so has practical ramifications since any physical implementation of an analog algorithm is expected to have some fixed precision. 

In this study, we examine the robustness of the QAUS algorithm to finite precision as exhibited by several noise models~\cite{RCNoise,PhysRevA.92.062320}. For completeness we also revisit the thermal robustness of the algorithm~\cite{wild,PhysRevA.79.022107,PhysRevA.75.062313,PhysRevA.71.060312,PhysRevA.72.042317,de_Vega_2010,PhysRevA.81.032305} using a specific decoherence model.  While these forms of imperfection are expected to appear together in any physical implementation, we treat each type separately here.  
We find that the quadratic speedup of the QAUS algorithm is sensitive to both finite precision and thermal effects, requiring both precision and temperature to scale in physically unreasonable ways to maintain the quantum speedup.  For the former, we do this using two forms of Hamiltonian implementation errors that shift the position of the minimum gap, and only with a precision that scales exponentially with the system size can the quadratic speedup be maintained.

 While it is well accepted that scalable quantum computing is not possible without fault tolerance \cite{PhysRevLett.96.050504}, there is as of yet no known accuracy-threshold theorem for the adiabatic paradigm of quantum computing.  Therefore, while fault tolerance schemes can be applied to the gate-based approach for solving unstructured search \cite{grover:97}, no equivalent schemes exist to date for the adiabatic approach.
Our study therefore calls into question the \emph{practical significance} of the QAUS asymptotic speedup in the absence of 
physically meaningful schemes to mitigate and correct for these errors. Specifically, if we are to rely on such speedups to give rise to a significant separation between the computational costs of quantum and classical algorithms at some maximum size,  there is a significant engineering challenge to realize the necessary quantum system with a sufficiently high precision, a feat that becomes increasingly harder with growing system size.

We begin with a brief overview of the algorithm and then move on to discuss the various types of imperfections considered and their impact on performance. In the concluding section we discuss the meaning and implications of our results.

\section{The quantum adiabatic unstructured search algorithm}

The unstructured search problem Hamiltonian is a one-dimensional projection onto the marked state:
\beq\label{eq:Hf}
H_p = \ident - |m\rangle\langle m| \, ,
\eeq
where $|m\rangle\langle m|$ is the projection onto the marked state, which belongs to the computational basis  $\{\ket{0}, \ket{1},\dots, \ket{N-1}\}$, and $ \ident $ is the identity operator.
To achieve the quadratic speedup, a carefully tailored variable-rate annealing schedule $s(t)$ is chosen that interpolates the Hamiltonian between a `beginning' Hamiltonian $H_b$ and the problem Hamiltonian $H_p$, varying slowly as a function of time $t \in [0,\mathcal{T}]$ in the vicinity of the minimum energy gap between the ground state and first excited state and more rapidly in places where the energy gap is large~\cite{jansen:07,lidarGap,kato:51,roland:02}. 
 Here, $H_b$ is a one-dimensional projection onto the equal superposition of all computational basis states, i.e., $H_b=\ident-|+\rangle \langle+|$, where $\ket{+} = \bigotimes_{i=1}^n \ket{+}_i$, and $\ket{+}_i = \frac{1}{\sqrt{2}} \left( \ket{0}_i + \ket{1}_i \right)$, and the total Hamiltonian is given by
\beq\label{eq:hs}
{H}(s(t))=(1-s(t)) H_b + s(t) H_p \,,
\eeq
where we have assumed the boundary conditions $s(0)=0$ and $s(\mathcal{T})=1$ at the beginning and end of the interpolation respectively. While the efficiency of a generic adiabatic algorithm may depend sensitively on the form of $H_b$ \cite{PhysRevA.73.022329,farhi:08}, the one-dimensional projection above gives rise to the optimal scaling performance~\cite{roland:02}.
 
If the initial state is taken to be the ground state of $H(0)$, i.e., $\ket{+}$, then the evolution according to $H(s)$ is restricted to the two-dimensional subspace spanned by $\ket{m}$ and $\ket{m^\perp} = \frac{1}{\sqrt{N-1}} \sum_{i \neq m}^N \ket{i}$.
 The ground state and first excited state of the system are in this subspace throughout the interpolation and can be written as:
\bes
\begin{align}
\ket{\veps_0(s)} & =  \cos \frac{\theta(s)}{2} \ket{m} + \sin \frac{\theta(s)}{2} \ket{m^{\perp}} \ , \\
\ket{\veps_1(s)}  &=  -\sin \frac{\theta(s)}{2} \ket{m} + \cos \frac{\theta(s)}{2} \ket{m^{\perp}} \ , 
\end{align}
\ees
with eigenvalues $\veps_0(s) = \frac{1}{2} \left( 1 - \delta(s) \right) \ , \quad \veps_1(s) =  \frac{1}{2} \left( 1 + \delta(s) \right)$ respectively and
\bes
\label{eq:gap-Grover}
\begin{align}
\delta(s) & =  \sqrt{ (1-2 s)^2 + \frac{4}{N} s (1-s) } \ , \label{eq:gap-Grover-delta} \\
\cos \theta(s) & =  \frac{1}{\delta(s)} \left[ 1 - 2 (1-s) \left( 1 - \frac{1}{N} \right) \right] \ , \label{eq:gap-Grover-cos} \\
\sin \theta(s) & =  \frac{2}{\delta(s)} \left(1 - s \right)  \frac{1}{\sqrt{N}}  \sqrt{1 - \frac{1}{N}} \ . \label{eq:gap-Grover-sin}
\end{align}
\ees
The remaining $N-2$ energy eigenstates are outside the aforementioned two-dimensional subspace and have energy $1$ throughout the interpolation. 
For later convenience, we write them as:
\bes
\begin{align}
&\ket{\veps_{k+1}(s)} = \frac{1}{\sqrt{2}}\left( \ket{f(k)} - \ket{\overline{f(k)}} \right) \ , \ 1 \leq k \leq \frac{N}{2} -1 \ , \\
&\ket{\veps_{2}'(s)}  =  \sqrt{\frac{N-2}{N-1}} \left(\ket{\overline{m}}  \right. \nonumber \\
& \hspace{2cm} \left.  -\frac{1}{N-2} \sum_{j=1}^{N/2-1} \left( \ket{f(j)} + \ket{\overline{f(j)}} \right)  \right) \ , \\
&\ket{\veps_{k+1}'(s)} = \sqrt{\frac{2(k-1)}{k}} \left(  \frac{1}{2} \left(\ket{f(k)} + \ket{\overline{f(k)}} \right)  \right. \nonumber \\
 &\quad \left.   - \frac{1}{2 (k-1)} \sum_{j=1}^{k-1} \left( \ket{f(j)} + \ket{\overline{f(j)}} \right) \right)  ,   2 \leq k \leq \frac{N}{2} -1 \ , \nonumber \\ & 
\end{align}
\ees
where $\overline{f(j)} = N - 1 - f(j)$ is the integer associated with the negation of the bit-representation of the integer $f(j)$ and
$$
f(j)=\left\{
  \begin{array}{@{}ll@{}}
    j-1, & \text{if}\ j-1 < \min(m,\overline{m}) \\
    j, & \text{otherwise}
  \end{array}\right. .
$$
This particular form of the excited states is useful because, $\sigma_i^z \left( \ket{f} + \ket{\overline{f}} \right) = \pm \left( \ket{f} - \ket{\overline{f}} \right)$ and $\sigma_i^z \left( \ket{f} - \ket{\overline{f}} \right) = \pm \left( \ket{f} + \ket{\overline{f}} \right)$, irrespective of the qubit index $i$ and the state $\ket{f}$.  This then means that we have the following relations:
\bes \label{eqt:SigmaZRelations}
\begin{align}
| \bra{ \veps_0(s) } \sigma_i^z \ket{\veps_{k+1}(s)} | & = \sqrt{\frac{2}{{N-1}}} \sin \frac{\theta(s)}{2}  \ , \nonumber \\
& \hspace{2cm} 1 \leq k \leq \frac{N}{2} -1 \ , \\
    | \bra{ \veps_0(s) } \sigma_i^z \ket{\veps'_2(s)} | & =  \frac{\sqrt{N-2}}{N-1} \sin \frac{\theta(s)}{2} \ , \\
  | \bra{ \veps_0(s) } \sigma_i^z \ket{\veps'_{k+1}(s)} | & = 0  \ , 2 \leq k \leq \frac{N}{2} -1 \ .
\end{align}
\ees
The optimized annealing schedule of Roland and Cerf~\cite{roland:02} that defines the QAUS algorithm satisfies a `local' adiabatic condition~\cite{hen:14,hen:14b}:
 \beq \label{eq:dsdt}
 \frac{\rmd s}{\rmd t} =\epsilon \delta^2(s) \ ,
 \eeq 
 where $\epsilon$ is a small constant. 
The optimized annealing schedule satisfying the interpolation boundary conditions is given by
 \beq \label{eqt:RCSchedule}
s(t) = \frac{1}{2} + \frac{1}{2 \sqrt{N-1}} \tan \left[ \left(2\frac{t}{\mathcal{T}}-1\right) \tan^{-1} \sqrt{N-1}  \right] \ ,
 \eeq
with  the optimal runtime being~\cite{roland:02}
\beq \label{eqt:AnnealingTime}
\mathcal{T} = \frac{N}{\epsilon \sqrt{N-1}} \tan^{-1} \sqrt{N-1} \approx \frac{\pi}{2\epsilon} \sqrt{N}\,,
\eeq
i.e., it is proportional to the square root of the dimension of the Hilbert space, similarly to its gate-based counterpart~\cite{grover:97}.  For a sufficiently small $\epsilon$, this choice guarantees that a system prepared in the ground state of $H(0)$ remains close to the instantaneous ground state throughout the evolution using $H(s)$.

\section{Finite schedule precision}

The QAUS algorithm is analog in nature, in that it requires continuously varying the strengths of $H_b$ and $H_p$ throughout the evolution~\cite{roland:02,roland:03,hen:14,hen:14b}. 
For the local adiabatic interpolation, Eq.~\eqref{eq:dsdt}, the annealing schedule $s(t)$ changes exponentially slowly around the minimum gap, which is on the order of $1/\sqrt{N}$, in a region of width $1/\sqrt{N}$~\cite{childsGoldstone,realizableAQCsearch}.
In any conceivable physical setting however, we expect only a limited control over the interpolating schedule, and here we ask whether this restriction adversely affect the performance of the QAUS algorithm.  

\begin{figure}[htbp] 
   \centering
   \subfigure[]{\includegraphics[width=0.95\columnwidth]{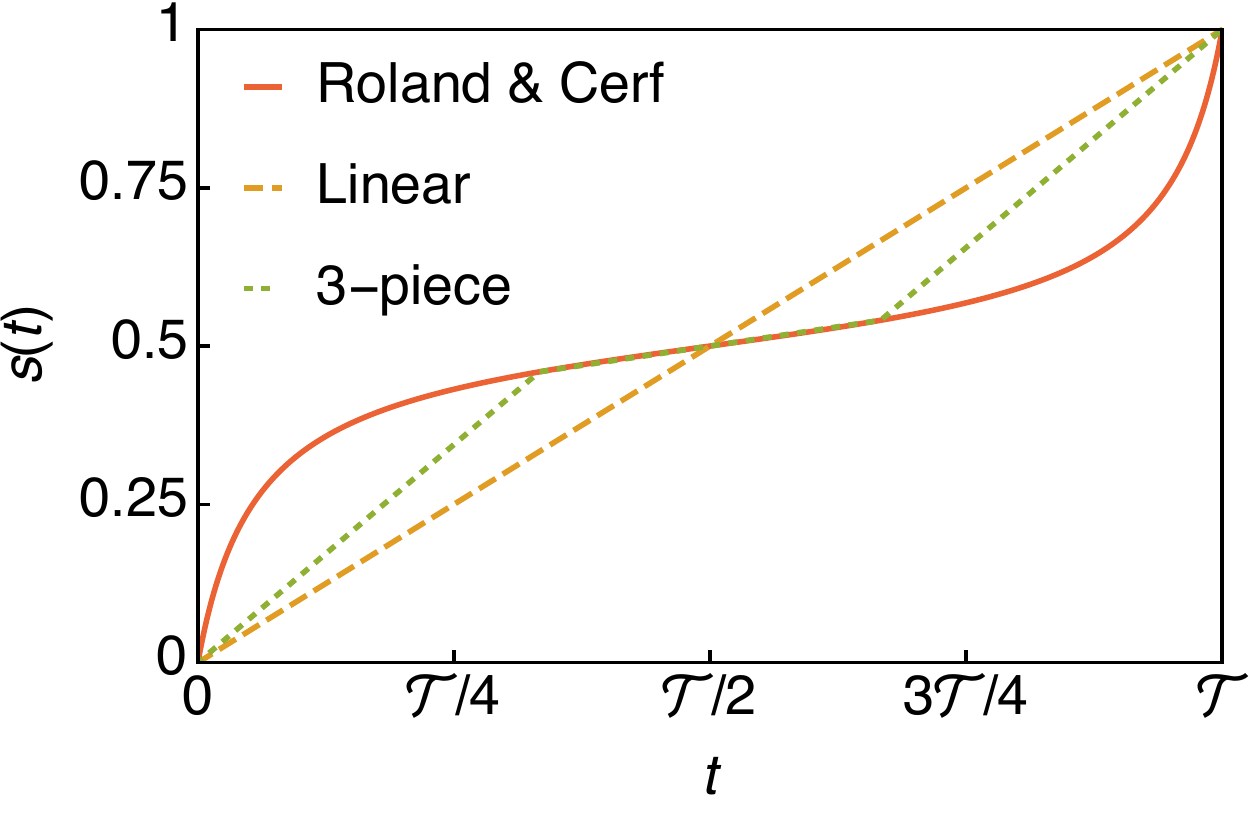} \label{fig:linSchedA}}
    \subfigure[]{\includegraphics[width=0.95\columnwidth]{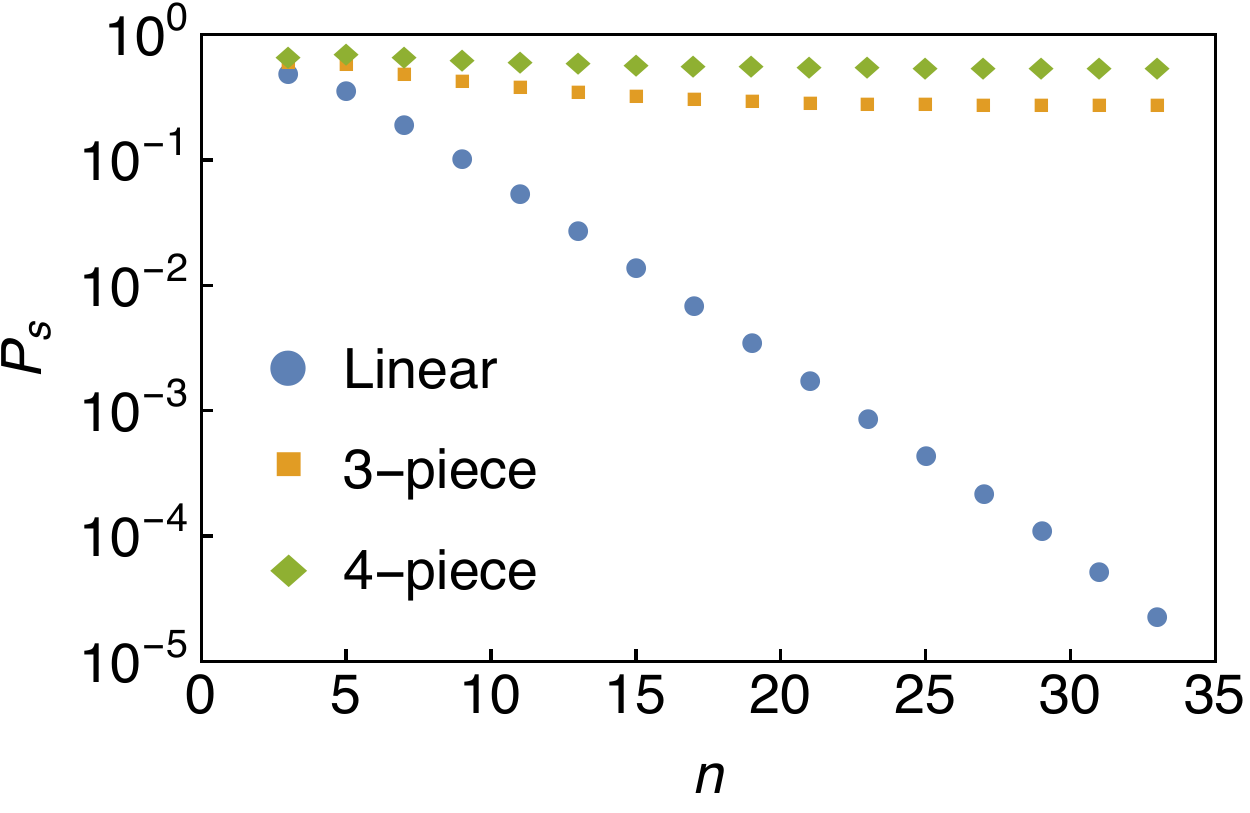} \label{fig:linSchedB}} 
   \caption{{ (a) Piecewise linear schedules that interpolate the QAUS schedule.} { (b) Probability of success $P_s$ as a function of problem size $n$ for several linearized schedules.  The simulations use $\mathcal{T}$ as given by Eq.~\eqref{eqt:AnnealingTime}, with $\epsilon = 0.01$.}}
   \label{fig:linSched}
\end{figure}

We begin our exploration by specifying the schedule $s(t)$ as a piecewise linear schedule between equally spaced time points $0, t_1, t_2,\ldots,\mathcal{T}$ with $t_j =j \Delta t$ for different spacings $\Delta t$ such that the schedule at $s(t_j)$ coincides with the original QAUS schedule given by Eq.~(\ref{eq:dsdt}) [see Fig.~\ref{fig:linSchedA}]. A numerical investigation reveals that a piecewise linear schedule with only two intermediate points (3-piece interpolation) suffices to achieve the quadratic speedup. This is demonstrated in Fig.~\ref{fig:linSchedB}, which depicts the probability of success $P_s$,  the probability of measuring the marked state at the end of the evolution, as a function of problem size $n$ for three- and four-piece interpolations.  The results show that already with a 3-piece schedule and a total time given by Eq.~\eqref{eqt:AnnealingTime}, a constant (with system size) probability of success is achieved. Higher-piece interpolations give, as expected, higher success probabilities.  

We thus find that the smooth $s(t)$ schedule in Eq.~\eqref{eqt:RCSchedule} is not necessary to obtain a quadratic speedup for as long as the linear slope at the minimum gap, $s = 1/2$, scales as $1/\sqrt{N}$. Since the region of the minimum gap shrinks exponentially as $1/\sqrt{N}$, this requires `hitting' the location of the minimum gap with increasing precision as the problem size grows.  

\begin{figure}[htbp] 
   \centering
   \includegraphics[width=0.95\columnwidth]{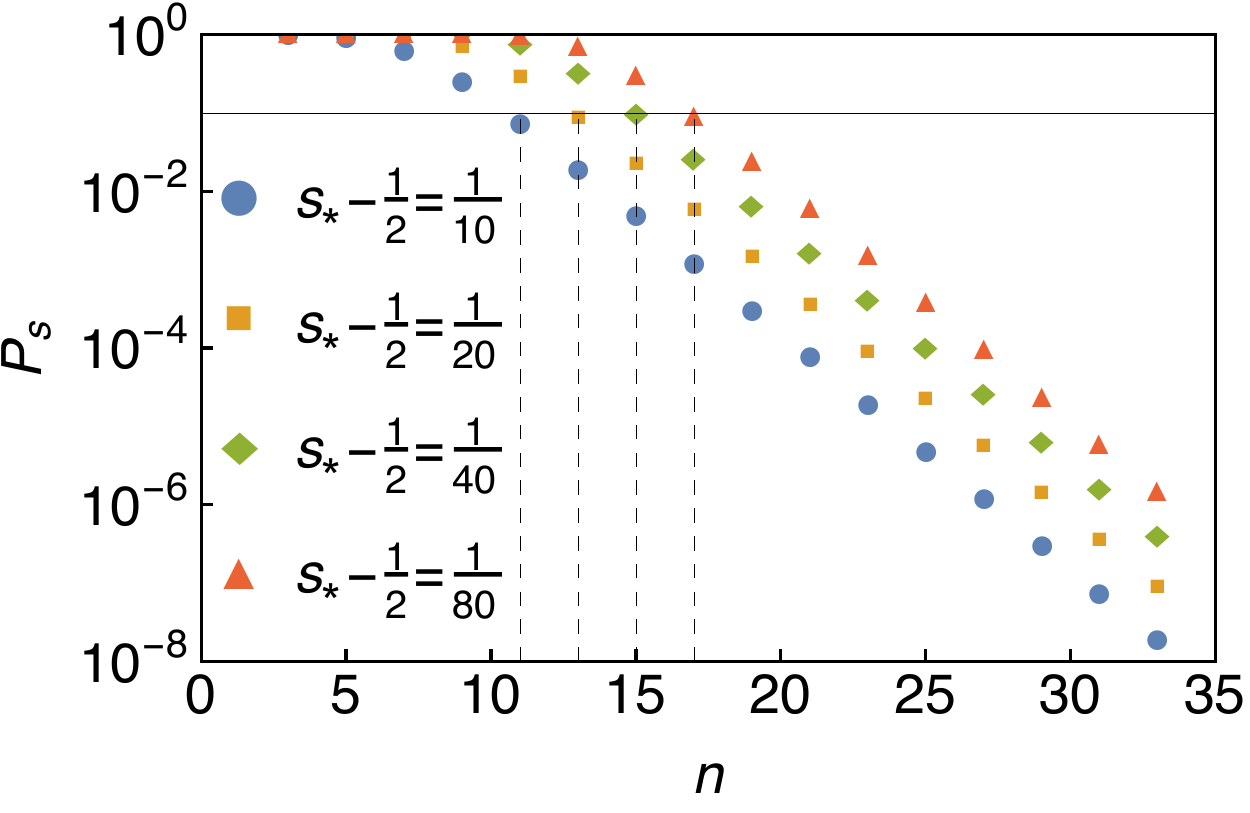} 
   \caption{{ Probability of success $P_s$ as a function of problem size $n$ (log-linear scale) for several values of $\chi$.} 
   Decay in success probability is exponential for any fixed value of $\chi$ for sufficiently large $n$. The simulations use the annealing schedule in Eq.~\eqref{eqt:RCSchedule} and $\mathcal{T}$ given by Eq.~\eqref{eqt:AnnealingTime}, with $\epsilon = 0.01$. The horizontal line indicates a fixed probability of $10^{-1}$, and the vertical dashed lines correspond to the maximum sizes $n=11,13,15,17$ before which the success probability drops below $10^{-1}$ for successive minimum gap positions $s_*$.}
   \label{fig:misH}
\end{figure}

To illustrate the above point, we consider the scenario of a slightly shifted schedule $s(t)$ that `misses' the location of the minimum gap by a small but fixed amount. This is equivalent to the case where the Hamiltonian itself is slightly misspecified:
\beq
H(s)=(1-s)H_b+s (1+\chi) H_p
\eeq
where $\chi$ is a small fixed constant and the schedule $s(t)$ is taken to be the unperturbed one [Eq.~\eqref{eqt:RCSchedule}]. For the above Hamiltonian, the gap is minimal at
\beq
s_*=\frac{N(\chi+2)-2(\chi+1)}{N(\chi+2)^2-4(\chi+1)} \ ,
\eeq
which, in the limit of $N \to \infty$ becomes $s_*=(\chi+2)^{-1}$. By employing the original QAUS annealing schedule, it is easy to see that there will always be a problem size $n_*$ beyond which the schedule does not sufficiently slow-down in the vicinity of the minimum gap. 
We confirm these expectations in Fig.~\ref{fig:misH} with simulation results for different values of displacements $\chi$ corresponding to displaced minimal gaps. Any nonzero value of $\chi$ (equivalently, any nonzero displacement of the minimum gap) eventually leads to an exponentially decreasing probability of success, with the transition to exponential behavior occurring at larger values of $n$ for smaller displacements $\chi$.  

To have the quadratic speedup, a fixed success probability must be maintained for growing system size.  The results in Fig.~\ref{fig:misH} show that to achieve this, the distance of $s_*$  from $1/2$ must be decreased accordingly. We can ask how big a perturbation is allowed, or equivalently how many bits of precision are required, for the schedule to achieve this. Since the gap is small to within a width of $1/\sqrt{N}$, we expect to require approximately $n/2$ bits of precision. Thus the schedule must be precise to within $O(n)$ bits of precision in order to maintain the quadratic speedup. This is  confirmed by the numerical data in Fig.~\ref{fig:misH}, where we see that for $n=11, 13, 15, 17$, we require approximately a factor of 2 decrease in the distance of $s_*$ from $1/2$. We further discuss the feasibility of the increasing precision requirement in the concluding section. 

\section{Noisy Hamiltonian}

The noise model in the previous section still restricted the unitary evolution 
to the two-dimensional subspace spanned by $\ket{m}$ and $\ket{m^\perp}$.  
We now extend our analysis by considering noise that prevents the evolution from being restricted to this subspace.  
Specially, we consider the QAUS algorithm perturbed by a noise Hamiltonian $\tilde{H}$ 
such that the total Hamiltonian is now given by $H'(s)=H(s)+\tilde{H}$, where the 
noise Hamiltonian $\tilde{H}$ has matrix elements in the $\ket{0}, \dots, \ket{N-1}$ basis 
that are drawn randomly from a Gaussian distribution with mean zero and standard deviation $\sigma$. 
Our model of noise has the elements fixed throughout the evolution, which is different from 
the time-dependent noise model studied in Ref.~\cite{RCNoise}. 
The adaptive step Runge-Kutta-Fehlberg algorithm was used 
for the efficient numerical solution of the time dependent Schr\"{o}dinger equation~\cite{press:92,cash:90}.

\begin{figure}[htbp]
\includegraphics[width=0.48\textwidth]{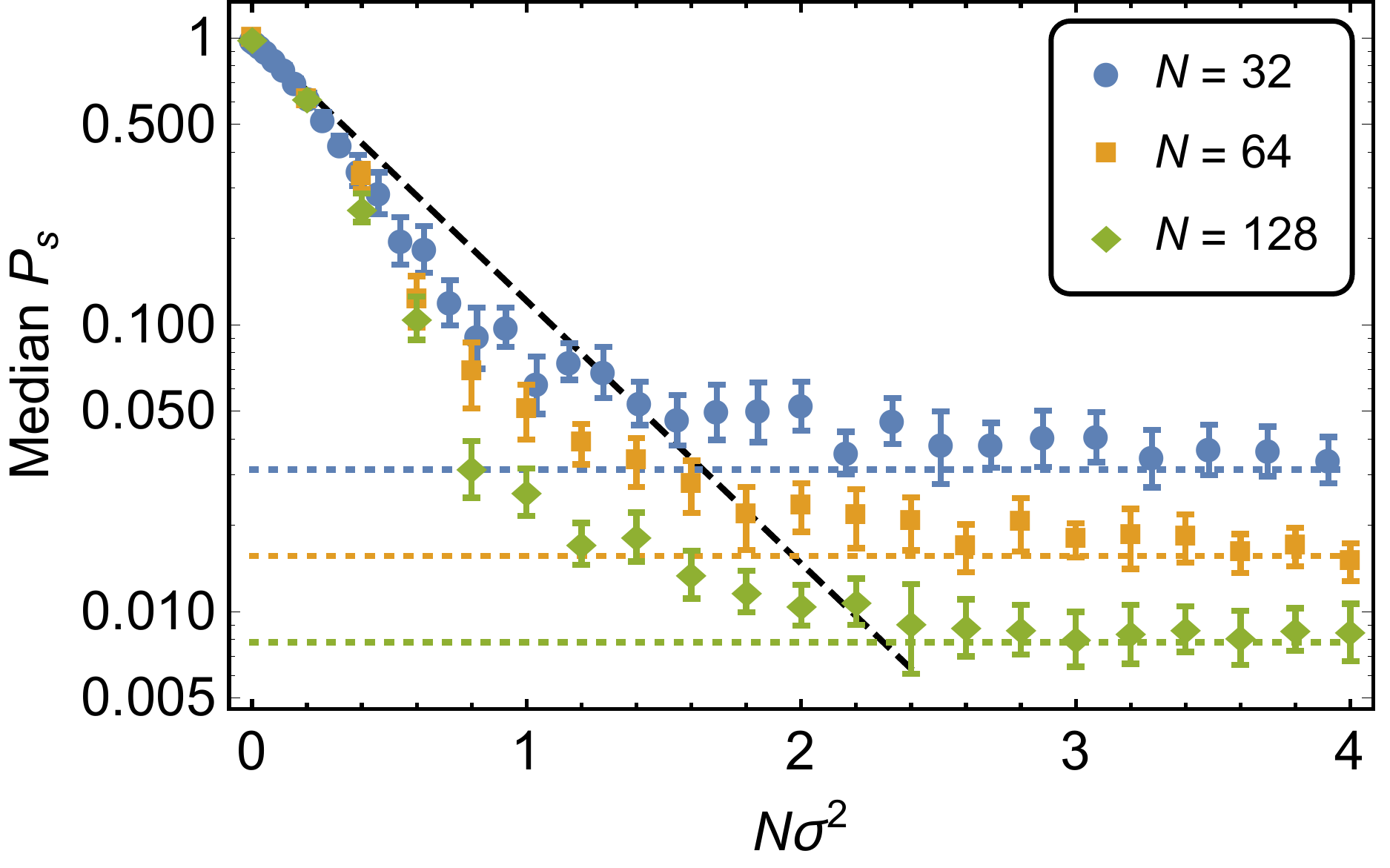}
\caption{The median probability of success $P_s$ as a function of noise strength $\sigma$ for different $N$.  For each $\sigma$ and $N$, 200 independent random error Hamiltonians were generated.  The data points and error bars are generated using $10^3$ bootstraps of the 200 runs, with the markers denoting the mean of the median and the error bars denoting two times the standard deviation of the median. The simulations use the annealing schedule in Eq.~\eqref{eqt:RCSchedule} and $\mathcal{T}$ given by Eq.~\eqref{eqt:AnnealingTime}, with $\epsilon = 0.01$. The diagonal dashed line corresponds to $P_s = \exp(-2.11 N \sigma^2)$. The horizontal dotted lines at $1/N$ are the asymptotic success probabilities for the various system sizes.}
\label{fig:noise}
\end{figure}

Figure~\ref{fig:noise} shows the dependence of the probability of success $P_s$ on $\sigma$ for various $N$ values.
The data can be fitted by
\beq
P_{s}\approx
\left\{
\begin{aligned}
\exp(-2.11N\sigma^2), & \qquad\mbox{for }\quad \sigma < \frac{1}{\sqrt{7N}}\\
\frac1{N}, & \qquad\mbox{for }\quad \sigma > \sqrt{\frac{3}{N}}
\end{aligned}
\right. .
\eeq
The probability of success approaches $1/N$ in the large noise limit.
In this limit the Hamiltonian is random so measuring the marked state occurs with probability $1/N$.
The initial exponential decay of $P_s$ is a function of $N \sigma^2$.  
This means that for a constant noise strength $\sigma$, the probability of success decays 
exponentially with $N$, and the only way to mitigate it is to require that $\sigma$, 
the noise strength, scales as $1/\sqrt{N}$.  

\section{Interaction with a thermal bath}
The robustness of the QAUS algorithm in the presence of interactions with an external environment has been extensively studied~\cite{wild,PhysRevA.79.022107,PhysRevA.75.062313,PhysRevA.71.060312,PhysRevA.72.042317,de_Vega_2010}.  A generic interaction breaks the symmetry that restricts the system evolution to the lowest two eigenstates, and for completeness here we show how the exponential number of excited states within a constant energy gap places (unrealistic) requirements on the temperature (or overall energy scale of the Hamiltonian) to maintain performance even for possibly the most innocuous noise model~\cite{PhysRevA.91.062320}. 

We consider a model of decoherence between a quantum annealing system of qubits and a thermal environment described by the Markovian adiabatic master equation~\cite{Albash_2012}. (We assume that this model holds throughout the anneal, even though we expect the validity conditions of the microscopic derivation of the model to break down near the minimum gap.)  We focus on the case where each qubit is connected to its own independent bath of bosonic harmonic oscillators.  The excitation rate from the ground state to an excited state $\ket{\veps_i(s)}$ at any point $s$ is generically given by $R_{0 \to i}(s) = \sum_{\alpha=1}^n \gamma(\Delta_i) e^{-\beta \Delta_i(s)} | \langle \veps_0(s) | A_\alpha | \veps_i(s) \rangle |^2$, where $\Delta_i(s)$ is the energy gap from the ground state to the excited state and $\beta$ is the inverse-temperature of the bosonic bath. $\gamma(\Delta_i)$ encodes the spectral density of the bosonic bath, the bath correlations, and the system-bath coupling strength $g$, and $A_\alpha$ is the system operator part of the system-bath interaction. We consider $A_\alpha = \sigma_\alpha^z$ corresponding to a `dephasing' bath.  For concreteness, we can consider an Ohmic spectral density, such that $\gamma(\Delta) = 2 \pi g^2  \Delta / (1 - e^{- \beta \Delta})$ \cite{Albash_2012}. 

Of relevance to us is the excitation rate during the anneal from the instantaneous ground state to all the excited states outside the two-dimensional subspace, which is given by
\begin{eqnarray}
R(s) & =& \sum_{i=2}^N R_{0\to i} = n  \gamma \left(\frac{1}{2}( 1+ \delta(s)) \right) e^{-\beta(1+\delta(s))/2}  \nonumber \\
&& \times \left( \left(\frac{N}{2}-1\right) | \langle \veps_{0}(s) | \sigma_1^z | \veps_2(s) \rangle |^2  \right. \nonumber \\
&& \left.  +  | \langle \veps_{0}(s) | \sigma_1^z | \veps_2'(s) \rangle |^2 \right) \ ,
\end{eqnarray}
where we have used the relations in Eq.~\eqref{eqt:SigmaZRelations}.

If follows from Eqs.~(\ref{eq:gap-Grover}) that $\sin^2\frac{\theta}2$ is a monotonically decreasing function of $s$ for $0 \leq s \leq \frac12$ and $N>1$, such that $\sin^2\frac{\theta}2 = 1 - \frac1{N}$ for $s=0$ and $\sin^2\frac{\theta}2 = \frac12 - \frac1{2\sqrt{N}}$ for $s=\frac12$.
Hence, we expect that the excitation rate to the excited states outside the two-dimensional subspace for the first half of the anneal to scale as $\sim n g^2 / (e^{\beta(1+\delta)/2}-1)$ for large $n$. In conjunction with a total annealing time that scales as $\sqrt{N}$ (Eq.~\eqref{eqt:AnnealingTime}),
we can expect the open-system dynamics to not be restricted to the two-dimensional subspace for a constant temperature and system-bath coupling.

We note that if the system thermalizes on the instantaneous Hamiltonian, the probability of success at any point in the anneal is given by
\beq
P_{s} \big|_{\mathrm{thermal}}= \frac{1}{1 + e^{- \beta \delta} + (N-2)e^{-\beta(1 + \delta)/2}} \ .
\eeq
For any fixed nonzero temperature, this gives a probability of being in the instantaneous ground state that scales as $1/N$ for any point along the interpolation.

In order to suppress the excitations to outside the two-dimensional subspace during the anneal, we can scale the inverse temperature $\beta$ linearly with $n$ for a constant $g$, which will exponentially (in $n$) suppress thermal excitations out of the ground state and will ensure that the instantaneous thermal state always has a finite population on the instantaneous ground state.  These results are consistent with the analysis of Ref.~\cite{wild}, although in that work the overall energy scale of the Hamiltonian $E_0$ was scaled linearly, such that $\beta E_0 \sim n$. 
Alternatively the system-bath coupling $g$ can be scaled down at least as $N^{-1/4}$ for a constant $\beta$ in order to ensure that thermal excitations are suppressed during the entire evolution.  

\section{Conclusions and discussion}

We studied the robustness of the quantum adiabatic unstructured search algorithm against various types of imperfections from limited control over the adiabatic schedule to Hamiltonian misspecification to an interaction with a decohering bath. Our findings can be summarized as follows. In the presence of finite perturbations to the Hamiltonian, the probability of hitting the marked state decreases exponentially with system size $n$  if the interpolating schedule is not adjusted accordingly.  This results in the loss of the quadratic speedup of the error-free algorithm. The scaling is similar when we consider a noise model that introduces Gaussian noise to the matrix elements, which now does not restrict the evolution to a two-dimensional subspace: the probability of hitting the marked state now decreases exponentially in $N$ for a fixed standard deviation of the noise. Our results indicate that the standard deviation must be scaled down as $1/\sqrt{N}$, which can also be derived from the analysis of Ref.~\cite{Jarret2018}\footnote{We thank Michael Jarret for pointing this out.}. While neither of the above noise models have been constructed with a physical mechanism in mind, these noise models reproduce effects we expect to generically occur. We expect generic noise to break the symmetries of the Hamiltonian that restrict the evolution to a particular subspace, and we expect generic noise to shift the position of the minimum gap in a noise-instance dependent way.  Our results show that without the interpolation schedule slowing down precisely at the noise-shifted minimum gap, the quadratic speedup of the QAUS algorithm will be lost.

We emphasize that even if the Hamiltonians $H_b$ and $H_p$ can be implemented precisely, the annealing schedule $s(t)$ still needs to be controlled with exponential precision around the minimum gap, even if we use piece-wise linear interpolations. This need for growing precision must inevitably translate to the need of additional resources, without which the QAUS algorithm cannot retain its quadratic speedup. This is the signature of analog computing, and our results illustrate the need for alternative methods that would combat the exponentially growing precision requirement.

Our work also has some implications for algorithms that require access to continuous-query Hamiltonian oracles or query other properties of the Hamiltonian (e.g., Refs.~\cite{Jarret2018,Cleve:2009:EDS:1536414.1536471}) wherein a sub-routine returning, e.g., the value of the gap, is called.  Our work suggests that the value of the gap needs to be returned with growing precision as a function of system size and hence requires growing space resources that needs to be accounted for.

We also revisited the thermal stability of the algorithm, studying it in the framework of the weak-coupling Markovian adiabatic master equation.  Here, in the absence of specific fine tuning, the presence of an exponential number of excited states at a fixed energy gap away from the ground state already imposes serious constraints on the temperature and/or the system-bath coupling just to ensure the evolution is restricted to the two-dimensional subspace.  The former needs to be scaled down at least inversely proportional to the system size, or the latter must be scaled down at least exponentially with the system size.

We finally point out that our analyses above is an asymptotic one. Since any physical device will have a finite fixed size, one could imagine noise strengths that are sufficiently reduced to make the QAUS algorithm successful. Such a device may still have practical uses if the computational costs of the quantum and classical algorithms are well seperated, and our results do not exclude such a possibility.

\begin{acknowledgements}
We thank Michael Jarret for useful discussions. 
The research is based upon work (partially) supported by the Office of
the Director of National Intelligence (ODNI), Intelligence Advanced
Research Projects Activity (IARPA), via the U.S. Army Research Office
contract W911NF-17-C-0050. The views and conclusions contained herein are
those of the authors and should not be interpreted as necessarily
representing the official policies or endorsements, either expressed or
implied, of the ODNI, IARPA, or the U.S. Government. The U.S. Government
is authorized to reproduce and distribute reprints for Governmental
purposes notwithstanding any copyright annotation thereon.
L. B. was supported within the framework of State Assignment 
No. 0033-2019-0007 of Russian Ministry of Science and Higher Education.
\end{acknowledgements}

\begin{thebibliography}{63}%
\makeatletter
\providecommand \@ifxundefined [1]{%
 \@ifx{#1\undefined}
}%
\providecommand \@ifnum [1]{%
 \ifnum #1\expandafter \@firstoftwo
 \else \expandafter \@secondoftwo
 \fi
}%
\providecommand \@ifx [1]{%
 \ifx #1\expandafter \@firstoftwo
 \else \expandafter \@secondoftwo
 \fi
}%
\providecommand \natexlab [1]{#1}%
\providecommand \enquote  [1]{``#1''}%
\providecommand \bibnamefont  [1]{#1}%
\providecommand \bibfnamefont [1]{#1}%
\providecommand \citenamefont [1]{#1}%
\providecommand \href@noop [0]{\@secondoftwo}%
\providecommand \href [0]{\begingroup \@sanitize@url \@href}%
\providecommand \@href[1]{\@@startlink{#1}\@@href}%
\providecommand \@@href[1]{\endgroup#1\@@endlink}%
\providecommand \@sanitize@url [0]{\catcode `\\12\catcode `\$12\catcode
  `\&12\catcode `\#12\catcode `\^12\catcode `\_12\catcode `\%12\relax}%
\providecommand \@@startlink[1]{}%
\providecommand \@@endlink[0]{}%
\providecommand \url  [0]{\begingroup\@sanitize@url \@url }%
\providecommand \@url [1]{\endgroup\@href {#1}{\urlprefix }}%
\providecommand \urlprefix  [0]{URL }%
\providecommand \Eprint [0]{\href }%
\providecommand \doibase [0]{http://dx.doi.org/}%
\providecommand \selectlanguage [0]{\@gobble}%
\providecommand \bibinfo  [0]{\@secondoftwo}%
\providecommand \bibfield  [0]{\@secondoftwo}%
\providecommand \translation [1]{[#1]}%
\providecommand \BibitemOpen [0]{}%
\providecommand \bibitemStop [0]{}%
\providecommand \bibitemNoStop [0]{.\EOS\space}%
\providecommand \EOS [0]{\spacefactor3000\relax}%
\providecommand \BibitemShut  [1]{\csname bibitem#1\endcsname}%
\let\auto@bib@innerbib\@empty
\bibitem [{\citenamefont {Finnila}\ \emph {et~al.}(1994)\citenamefont
  {Finnila}, \citenamefont {Gomez}, \citenamefont {Sebenik}, \citenamefont
  {Stenson},\ and\ \citenamefont {Doll}}]{Finnila1994343}%
  \BibitemOpen
  \bibfield  {author} {\bibinfo {author} {\bibfnamefont {A.B.}\ \bibnamefont
  {Finnila}}, \bibinfo {author} {\bibfnamefont {M.A.}\ \bibnamefont {Gomez}},
  \bibinfo {author} {\bibfnamefont {C.}~\bibnamefont {Sebenik}}, \bibinfo
  {author} {\bibfnamefont {C.}~\bibnamefont {Stenson}}, \ and\ \bibinfo
  {author} {\bibfnamefont {J.D.}\ \bibnamefont {Doll}},\ }\bibfield  {title}
  {\enquote {\bibinfo {title} {Quantum annealing: A new method for minimizing
  multidimensional functions},}\ }\href {\doibase
  http://dx.doi.org/10.1016/0009-2614(94)00117-0} {\bibfield  {journal}
  {\bibinfo  {journal} {Chemical Physics Letters}\ }\textbf {\bibinfo {volume}
  {219}},\ \bibinfo {pages} {343 -- 348} (\bibinfo {year} {1994})}\BibitemShut
  {NoStop}%
\bibitem [{\citenamefont {Brooke}\ \emph {et~al.}(1999)\citenamefont {Brooke},
  \citenamefont {Bitko}, \citenamefont {F.}, \citenamefont {Rosenbaum},\ and\
  \citenamefont {Aeppli}}]{Brooke30041999}%
  \BibitemOpen
  \bibfield  {author} {\bibinfo {author} {\bibfnamefont {J.}~\bibnamefont
  {Brooke}}, \bibinfo {author} {\bibfnamefont {D.}~\bibnamefont {Bitko}},
  \bibinfo {author} {\bibfnamefont {T.}~\bibnamefont {F.}}, \bibinfo {author}
  {\bibnamefont {Rosenbaum}}, \ and\ \bibinfo {author} {\bibfnamefont
  {G.}~\bibnamefont {Aeppli}},\ }\bibfield  {title} {\enquote {\bibinfo {title}
  {Quantum annealing of a disordered magnet},}\ }\href {\doibase
  10.1126/science.284.5415.779} {\bibfield  {journal} {\bibinfo  {journal}
  {Science}\ }\textbf {\bibinfo {volume} {284}},\ \bibinfo {pages} {779--781}
  (\bibinfo {year} {1999})}\BibitemShut {NoStop}%
\bibitem [{\citenamefont {Kadowaki}\ and\ \citenamefont
  {Nishimori}(1998)}]{kadowaki:98}%
  \BibitemOpen
  \bibfield  {author} {\bibinfo {author} {\bibfnamefont {Tadashi}\ \bibnamefont
  {Kadowaki}}\ and\ \bibinfo {author} {\bibfnamefont {Hidetoshi}\ \bibnamefont
  {Nishimori}},\ }\bibfield  {title} {\enquote {\bibinfo {title} {Quantum
  annealing in the transverse ising model},}\ }\href {\doibase
  10.1103/PhysRevE.58.5355} {\bibfield  {journal} {\bibinfo  {journal} {Phys.
  Rev. E}\ }\textbf {\bibinfo {volume} {58}},\ \bibinfo {pages} {5355--5363}
  (\bibinfo {year} {1998})}\BibitemShut {NoStop}%
\bibitem [{\citenamefont {Farhi}\ \emph {et~al.}(2001)\citenamefont {Farhi},
  \citenamefont {Goldstone}, \citenamefont {Gutmann}, \citenamefont {Lapan},
  \citenamefont {Lundgren},\ and\ \citenamefont {Preda}}]{farhi:01}%
  \BibitemOpen
  \bibfield  {author} {\bibinfo {author} {\bibfnamefont {Edward}\ \bibnamefont
  {Farhi}}, \bibinfo {author} {\bibfnamefont {Jeffrey}\ \bibnamefont
  {Goldstone}}, \bibinfo {author} {\bibfnamefont {Sam}\ \bibnamefont
  {Gutmann}}, \bibinfo {author} {\bibfnamefont {Joshua}\ \bibnamefont {Lapan}},
  \bibinfo {author} {\bibfnamefont {Andrew}\ \bibnamefont {Lundgren}}, \ and\
  \bibinfo {author} {\bibfnamefont {Daniel}\ \bibnamefont {Preda}},\ }\bibfield
   {title} {\enquote {\bibinfo {title} {A quantum adiabatic evolution algorithm
  applied to random instances of an np-complete problem},}\ }\href {\doibase
  10.1126/science.1057726} {\bibfield  {journal} {\bibinfo  {journal}
  {Science}\ }\textbf {\bibinfo {volume} {292}},\ \bibinfo {pages} {472--475}
  (\bibinfo {year} {2001})}\BibitemShut {NoStop}%
\bibitem [{\citenamefont {Santoro}\ \emph {et~al.}(2002)\citenamefont
  {Santoro}, \citenamefont {Marto{\v n}{\'a}k}, \citenamefont {Tosatti},\ and\
  \citenamefont {Car}}]{santoro:02}%
  \BibitemOpen
  \bibfield  {author} {\bibinfo {author} {\bibfnamefont {Giuseppe~E.}\
  \bibnamefont {Santoro}}, \bibinfo {author} {\bibfnamefont {Roman}\
  \bibnamefont {Marto{\v n}{\'a}k}}, \bibinfo {author} {\bibfnamefont {Erio}\
  \bibnamefont {Tosatti}}, \ and\ \bibinfo {author} {\bibfnamefont {Roberto}\
  \bibnamefont {Car}},\ }\bibfield  {title} {\enquote {\bibinfo {title} {Theory
  of quantum annealing of an ising spin glass},}\ }\href {\doibase
  10.1126/science.1068774} {\bibfield  {journal} {\bibinfo  {journal}
  {Science}\ }\textbf {\bibinfo {volume} {295}},\ \bibinfo {pages} {2427--2430}
  (\bibinfo {year} {2002})}\BibitemShut {NoStop}%
\bibitem [{\citenamefont {van Dam}\ \emph {et~al.}(2001)\citenamefont {van
  Dam}, \citenamefont {Mosca},\ and\ \citenamefont {Vazirani}}]{howPowerful}%
  \BibitemOpen
  \bibfield  {author} {\bibinfo {author} {\bibfnamefont {W.}~\bibnamefont {van
  Dam}}, \bibinfo {author} {\bibfnamefont {M.}~\bibnamefont {Mosca}}, \ and\
  \bibinfo {author} {\bibfnamefont {U.}~\bibnamefont {Vazirani}},\ }\bibfield
  {title} {\enquote {\bibinfo {title} {How powerful is adiabatic quantum
  computation?}}\ }in\ \href {\doibase 10.1109/SFCS.2001.959902} {\emph
  {\bibinfo {booktitle} {Proceedings 2001 IEEE International Conference on
  Cluster Computing}}}\ (\bibinfo {year} {2001})\ pp.\ \bibinfo {pages}
  {279--287}\BibitemShut {NoStop}%
\bibitem [{\citenamefont
  {Reichardt}(2004)}]{Reichardt:2004:QAO:1007352.1007428}%
  \BibitemOpen
  \bibfield  {author} {\bibinfo {author} {\bibfnamefont {Ben~W.}\ \bibnamefont
  {Reichardt}},\ }\bibfield  {title} {\enquote {\bibinfo {title} {The quantum
  adiabatic optimization algorithm and local minima},}\ }in\ \href {\doibase
  10.1145/1007352.1007428} {\emph {\bibinfo {booktitle} {Proceedings of the
  Thirty-sixth Annual ACM Symposium on Theory of Computing}}},\ \bibinfo
  {series and number} {STOC '04}\ (\bibinfo  {publisher} {ACM},\ \bibinfo
  {address} {New York, NY, USA},\ \bibinfo {year} {2004})\ pp.\ \bibinfo
  {pages} {502--510}\BibitemShut {NoStop}%
\bibitem [{\citenamefont {Young}\ \emph {et~al.}(2008)\citenamefont {Young},
  \citenamefont {Knysh},\ and\ \citenamefont {Smelyanskiy}}]{young:08}%
  \BibitemOpen
  \bibfield  {author} {\bibinfo {author} {\bibfnamefont {A.~P.}\ \bibnamefont
  {Young}}, \bibinfo {author} {\bibfnamefont {S.}~\bibnamefont {Knysh}}, \ and\
  \bibinfo {author} {\bibfnamefont {V.~N.}\ \bibnamefont {Smelyanskiy}},\
  }\bibfield  {title} {\enquote {\bibinfo {title} {Size dependence of the
  minimum excitation gap in the quantum adiabatic algorithm},}\ }\href
  {\doibase 10.1103/PhysRevLett.101.170503} {\bibfield  {journal} {\bibinfo
  {journal} {Phys. Rev. Lett.}\ }\textbf {\bibinfo {volume} {101}},\ \bibinfo
  {pages} {170503} (\bibinfo {year} {2008})}\BibitemShut {NoStop}%
\bibitem [{\citenamefont {Young}\ \emph {et~al.}(2010)\citenamefont {Young},
  \citenamefont {Knysh},\ and\ \citenamefont {Smelyanskiy}}]{young:10}%
  \BibitemOpen
  \bibfield  {author} {\bibinfo {author} {\bibfnamefont {A.~P.}\ \bibnamefont
  {Young}}, \bibinfo {author} {\bibfnamefont {S.}~\bibnamefont {Knysh}}, \ and\
  \bibinfo {author} {\bibfnamefont {V.~N.}\ \bibnamefont {Smelyanskiy}},\
  }\bibfield  {title} {\enquote {\bibinfo {title} {First-order phase transition
  in the quantum adiabatic algorithm},}\ }\href {\doibase
  10.1103/PhysRevLett.104.020502} {\bibfield  {journal} {\bibinfo  {journal}
  {Phys. Rev. Lett.}\ }\textbf {\bibinfo {volume} {104}},\ \bibinfo {pages}
  {020502} (\bibinfo {year} {2010})}\BibitemShut {NoStop}%
\bibitem [{\citenamefont {Hen}\ and\ \citenamefont {Young}(2011)}]{hen:11}%
  \BibitemOpen
  \bibfield  {author} {\bibinfo {author} {\bibfnamefont {Itay}\ \bibnamefont
  {Hen}}\ and\ \bibinfo {author} {\bibfnamefont {A.~P.}\ \bibnamefont
  {Young}},\ }\bibfield  {title} {\enquote {\bibinfo {title} {Exponential
  complexity of the quantum adiabatic algorithm for certain satisfiability
  problems},}\ }\href {\doibase 10.1103/PhysRevE.84.061152} {\bibfield
  {journal} {\bibinfo  {journal} {Phys. Rev. E}\ }\textbf {\bibinfo {volume}
  {84}},\ \bibinfo {pages} {061152} (\bibinfo {year} {2011})}\BibitemShut
  {NoStop}%
\bibitem [{\citenamefont {Hen}(2012)}]{hen:12}%
  \BibitemOpen
  \bibfield  {author} {\bibinfo {author} {\bibfnamefont {Itay}\ \bibnamefont
  {Hen}},\ }\bibfield  {title} {\enquote {\bibinfo {title} {Excitation gap from
  optimized correlation functions in quantum monte carlo simulations},}\ }\href
  {\doibase 10.1103/PhysRevE.85.036705} {\bibfield  {journal} {\bibinfo
  {journal} {Phys. Rev. E}\ }\textbf {\bibinfo {volume} {85}},\ \bibinfo
  {pages} {036705} (\bibinfo {year} {2012})}\BibitemShut {NoStop}%
\bibitem [{\citenamefont {Farhi}\ \emph {et~al.}(2012)\citenamefont {Farhi},
  \citenamefont {Gosset}, \citenamefont {Hen}, \citenamefont {Sandvik},
  \citenamefont {Shor}, \citenamefont {Young},\ and\ \citenamefont
  {Zamponi}}]{farhi:12}%
  \BibitemOpen
  \bibfield  {author} {\bibinfo {author} {\bibfnamefont {Edward}\ \bibnamefont
  {Farhi}}, \bibinfo {author} {\bibfnamefont {David}\ \bibnamefont {Gosset}},
  \bibinfo {author} {\bibfnamefont {Itay}\ \bibnamefont {Hen}}, \bibinfo
  {author} {\bibfnamefont {A.~W.}\ \bibnamefont {Sandvik}}, \bibinfo {author}
  {\bibfnamefont {Peter}\ \bibnamefont {Shor}}, \bibinfo {author}
  {\bibfnamefont {A.~P.}\ \bibnamefont {Young}}, \ and\ \bibinfo {author}
  {\bibfnamefont {Francesco}\ \bibnamefont {Zamponi}},\ }\bibfield  {title}
  {\enquote {\bibinfo {title} {Performance of the quantum adiabatic algorithm
  on random instances of two optimization problems on regular hypergraphs},}\
  }\href {\doibase 10.1103/PhysRevA.86.052334} {\bibfield  {journal} {\bibinfo
  {journal} {Phys. Rev. A}\ }\textbf {\bibinfo {volume} {86}},\ \bibinfo
  {pages} {052334} (\bibinfo {year} {2012})}\BibitemShut {NoStop}%
\bibitem [{\citenamefont {Sch\"utzhold}\ and\ \citenamefont
  {Schaller}(2006)}]{PhysRevA.74.060304}%
  \BibitemOpen
  \bibfield  {author} {\bibinfo {author} {\bibfnamefont {Ralf}\ \bibnamefont
  {Sch\"utzhold}}\ and\ \bibinfo {author} {\bibfnamefont {Gernot}\ \bibnamefont
  {Schaller}},\ }\bibfield  {title} {\enquote {\bibinfo {title} {Adiabatic
  quantum algorithms as quantum phase transitions: First versus second
  order},}\ }\href {\doibase 10.1103/PhysRevA.74.060304} {\bibfield  {journal}
  {\bibinfo  {journal} {Phys. Rev. A}\ }\textbf {\bibinfo {volume} {74}},\
  \bibinfo {pages} {060304} (\bibinfo {year} {2006})}\BibitemShut {NoStop}%
\bibitem [{\citenamefont {\ifmmode \check{Z}\else
  \v{Z}\fi{}nidari\ifmmode~\check{c}\else \v{c}\fi{}}\ and\ \citenamefont
  {Horvat}(2006)}]{PhysRevA.73.022329}%
  \BibitemOpen
  \bibfield  {author} {\bibinfo {author} {\bibfnamefont {Marko}\ \bibnamefont
  {\ifmmode \check{Z}\else \v{Z}\fi{}nidari\ifmmode~\check{c}\else
  \v{c}\fi{}}}\ and\ \bibinfo {author} {\bibfnamefont {Martin}\ \bibnamefont
  {Horvat}},\ }\bibfield  {title} {\enquote {\bibinfo {title} {Exponential
  complexity of an adiabatic algorithm for an np-complete problem},}\ }\href
  {\doibase 10.1103/PhysRevA.73.022329} {\bibfield  {journal} {\bibinfo
  {journal} {Phys. Rev. A}\ }\textbf {\bibinfo {volume} {73}},\ \bibinfo
  {pages} {022329} (\bibinfo {year} {2006})}\BibitemShut {NoStop}%
\bibitem [{\citenamefont {Vandersypen}\ \emph {et~al.}(2001)\citenamefont
  {Vandersypen}, \citenamefont {Steffen}, \citenamefont {Breyta}, \citenamefont
  {Yannoni}, \citenamefont {Sherwood},\ and\ \citenamefont
  {Chuang}}]{vandersypen:01}%
  \BibitemOpen
  \bibfield  {author} {\bibinfo {author} {\bibfnamefont {Lieven M.~K.}\
  \bibnamefont {Vandersypen}}, \bibinfo {author} {\bibfnamefont {Matthias}\
  \bibnamefont {Steffen}}, \bibinfo {author} {\bibfnamefont {Gregory}\
  \bibnamefont {Breyta}}, \bibinfo {author} {\bibfnamefont {Costantino~S.}\
  \bibnamefont {Yannoni}}, \bibinfo {author} {\bibfnamefont {Mark~H.}\
  \bibnamefont {Sherwood}}, \ and\ \bibinfo {author} {\bibfnamefont {Isaac~L.}\
  \bibnamefont {Chuang}},\ }\bibfield  {title} {\enquote {\bibinfo {title}
  {Experimental realization of shor's quantum factoring algorithm using nuclear
  magnetic resonance},}\ }\href {\doibase 10.1038/414883a} {\bibfield
  {journal} {\bibinfo  {journal} {Nature}\ }\textbf {\bibinfo {volume} {414}},\
  \bibinfo {pages} {883--887} (\bibinfo {year} {2001})}\BibitemShut {NoStop}%
\bibitem [{\citenamefont {Bian}\ \emph {et~al.}(2013)\citenamefont {Bian},
  \citenamefont {Chudak}, \citenamefont {Macready}, \citenamefont {Clark},\
  and\ \citenamefont {Gaitan}}]{gaitan:12}%
  \BibitemOpen
  \bibfield  {author} {\bibinfo {author} {\bibfnamefont {Zhengbing}\
  \bibnamefont {Bian}}, \bibinfo {author} {\bibfnamefont {Fabian}\ \bibnamefont
  {Chudak}}, \bibinfo {author} {\bibfnamefont {William~G.}\ \bibnamefont
  {Macready}}, \bibinfo {author} {\bibfnamefont {Lane}\ \bibnamefont {Clark}},
  \ and\ \bibinfo {author} {\bibfnamefont {Frank}\ \bibnamefont {Gaitan}},\
  }\bibfield  {title} {\enquote {\bibinfo {title} {Experimental determination
  of ramsey numbers},}\ }\href {\doibase 10.1103/PhysRevLett.111.130505}
  {\bibfield  {journal} {\bibinfo  {journal} {Phys. Rev. Lett.}\ }\textbf
  {\bibinfo {volume} {111}},\ \bibinfo {pages} {130505} (\bibinfo {year}
  {2013})}\BibitemShut {NoStop}%
\bibitem [{\citenamefont {Babbush}\ \emph {et~al.}(2014)\citenamefont
  {Babbush}, \citenamefont {Love},\ and\ \citenamefont
  {Aspuru-Guzik}}]{Babbush:2014}%
  \BibitemOpen
  \bibfield  {author} {\bibinfo {author} {\bibfnamefont {Ryan}\ \bibnamefont
  {Babbush}}, \bibinfo {author} {\bibfnamefont {Peter~J.}\ \bibnamefont
  {Love}}, \ and\ \bibinfo {author} {\bibfnamefont {Al{\'a}n}\ \bibnamefont
  {Aspuru-Guzik}},\ }\bibfield  {title} {\enquote {\bibinfo {title} {Adiabatic
  quantum simulation of quantum chemistry},}\ }\href
  {http://dx.doi.org/10.1038/srep06603} {\bibfield  {journal} {\bibinfo
  {journal} {Scientific Reports}\ }\textbf {\bibinfo {volume} {4}},\ \bibinfo
  {pages} {6603 EP --} (\bibinfo {year} {2014})}\BibitemShut {NoStop}%
\bibitem [{\citenamefont {R{\o}nnow}\ \emph {et~al.}(2014)\citenamefont
  {R{\o}nnow}, \citenamefont {Wang}, \citenamefont {Job}, \citenamefont
  {Boixo}, \citenamefont {Isakov}, \citenamefont {Wecker}, \citenamefont
  {Martinis}, \citenamefont {Lidar},\ and\ \citenamefont {Troyer}}]{speedup}%
  \BibitemOpen
  \bibfield  {author} {\bibinfo {author} {\bibfnamefont {Troels~F.}\
  \bibnamefont {R{\o}nnow}}, \bibinfo {author} {\bibfnamefont {Zhihui}\
  \bibnamefont {Wang}}, \bibinfo {author} {\bibfnamefont {Joshua}\ \bibnamefont
  {Job}}, \bibinfo {author} {\bibfnamefont {Sergio}\ \bibnamefont {Boixo}},
  \bibinfo {author} {\bibfnamefont {Sergei~V.}\ \bibnamefont {Isakov}},
  \bibinfo {author} {\bibfnamefont {David}\ \bibnamefont {Wecker}}, \bibinfo
  {author} {\bibfnamefont {John~M.}\ \bibnamefont {Martinis}}, \bibinfo
  {author} {\bibfnamefont {Daniel~A.}\ \bibnamefont {Lidar}}, \ and\ \bibinfo
  {author} {\bibfnamefont {Matthias}\ \bibnamefont {Troyer}},\ }\bibfield
  {title} {\enquote {\bibinfo {title} {Defining and detecting quantum
  speedup},}\ }\href {\doibase 10.1126/science.1252319} {\bibfield  {journal}
  {\bibinfo  {journal} {Science}\ }\textbf {\bibinfo {volume} {345}},\ \bibinfo
  {pages} {420--424} (\bibinfo {year} {2014})}\BibitemShut {NoStop}%
\bibitem [{\citenamefont {Johnson}\ \emph {et~al.}(2011)\citenamefont
  {Johnson}, \citenamefont {Amin}, \citenamefont {Gildert}, \citenamefont
  {Lanting}, \citenamefont {Hamze}, \citenamefont {Dickson}, \citenamefont
  {Harris}, \citenamefont {Berkley}, \citenamefont {Johansson}, \citenamefont
  {Bunyk}, \citenamefont {Chapple}, \citenamefont {Enderud}, \citenamefont
  {Hilton}, \citenamefont {Karimi}, \citenamefont {Ladizinsky}, \citenamefont
  {Ladizinsky}, \citenamefont {Oh}, \citenamefont {Perminov}, \citenamefont
  {Rich}, \citenamefont {Thom}, \citenamefont {Tolkacheva}, \citenamefont
  {Truncik}, \citenamefont {Uchaikin}, \citenamefont {Wang}, \citenamefont
  {Wilson},\ and\ \citenamefont {Rose}}]{DWave}%
  \BibitemOpen
  \bibfield  {author} {\bibinfo {author} {\bibfnamefont {M.~W.}\ \bibnamefont
  {Johnson}}, \bibinfo {author} {\bibfnamefont {M.~H.~S.}\ \bibnamefont
  {Amin}}, \bibinfo {author} {\bibfnamefont {S.}~\bibnamefont {Gildert}},
  \bibinfo {author} {\bibfnamefont {T.}~\bibnamefont {Lanting}}, \bibinfo
  {author} {\bibfnamefont {F.}~\bibnamefont {Hamze}}, \bibinfo {author}
  {\bibfnamefont {N.}~\bibnamefont {Dickson}}, \bibinfo {author} {\bibfnamefont
  {R.}~\bibnamefont {Harris}}, \bibinfo {author} {\bibfnamefont {A.~J.}\
  \bibnamefont {Berkley}}, \bibinfo {author} {\bibfnamefont {J.}~\bibnamefont
  {Johansson}}, \bibinfo {author} {\bibfnamefont {P.}~\bibnamefont {Bunyk}},
  \bibinfo {author} {\bibfnamefont {E.~M.}\ \bibnamefont {Chapple}}, \bibinfo
  {author} {\bibfnamefont {C.}~\bibnamefont {Enderud}}, \bibinfo {author}
  {\bibfnamefont {J.~P.}\ \bibnamefont {Hilton}}, \bibinfo {author}
  {\bibfnamefont {K.}~\bibnamefont {Karimi}}, \bibinfo {author} {\bibfnamefont
  {E.}~\bibnamefont {Ladizinsky}}, \bibinfo {author} {\bibfnamefont
  {N.}~\bibnamefont {Ladizinsky}}, \bibinfo {author} {\bibfnamefont
  {T.}~\bibnamefont {Oh}}, \bibinfo {author} {\bibfnamefont {I.}~\bibnamefont
  {Perminov}}, \bibinfo {author} {\bibfnamefont {C.}~\bibnamefont {Rich}},
  \bibinfo {author} {\bibfnamefont {M.~C.}\ \bibnamefont {Thom}}, \bibinfo
  {author} {\bibfnamefont {E.}~\bibnamefont {Tolkacheva}}, \bibinfo {author}
  {\bibfnamefont {C.~J.~S.}\ \bibnamefont {Truncik}}, \bibinfo {author}
  {\bibfnamefont {S.}~\bibnamefont {Uchaikin}}, \bibinfo {author}
  {\bibfnamefont {J.}~\bibnamefont {Wang}}, \bibinfo {author} {\bibfnamefont
  {B.}~\bibnamefont {Wilson}}, \ and\ \bibinfo {author} {\bibfnamefont
  {G.}~\bibnamefont {Rose}},\ }\bibfield  {title} {\enquote {\bibinfo {title}
  {Quantum annealing with manufactured spins},}\ }\href {\doibase
  10.1038/nature10012} {\bibfield  {journal} {\bibinfo  {journal} {Nature}\
  }\textbf {\bibinfo {volume} {473}},\ \bibinfo {pages} {194--198} (\bibinfo
  {year} {2011})}\BibitemShut {NoStop}%
\bibitem [{\citenamefont {Hen}\ \emph {et~al.}(2015)\citenamefont {Hen},
  \citenamefont {Job}, \citenamefont {Albash}, \citenamefont {R{\o}nnow},
  \citenamefont {Troyer},\ and\ \citenamefont {Lidar}}]{Hen:2015rt}%
  \BibitemOpen
  \bibfield  {author} {\bibinfo {author} {\bibfnamefont {Itay}\ \bibnamefont
  {Hen}}, \bibinfo {author} {\bibfnamefont {Joshua}\ \bibnamefont {Job}},
  \bibinfo {author} {\bibfnamefont {Tameem}\ \bibnamefont {Albash}}, \bibinfo
  {author} {\bibfnamefont {Troels~F.}\ \bibnamefont {R{\o}nnow}}, \bibinfo
  {author} {\bibfnamefont {Matthias}\ \bibnamefont {Troyer}}, \ and\ \bibinfo
  {author} {\bibfnamefont {Daniel~A.}\ \bibnamefont {Lidar}},\ }\bibfield
  {title} {\enquote {\bibinfo {title} {Probing for quantum speedup in
  spin-glass problems with planted solutions},}\ }\href
  {http://link.aps.org/doi/10.1103/PhysRevA.92.042325} {\bibfield  {journal}
  {\bibinfo  {journal} {{Phys. Rev. A}}\ }\textbf {\bibinfo {volume} {92}},\
  \bibinfo {pages} {042325--} (\bibinfo {year} {2015})}\BibitemShut {NoStop}%
\bibitem [{\citenamefont {Boixo}\ \emph {et~al.}(2013)\citenamefont {Boixo},
  \citenamefont {Albash}, \citenamefont {Spedalieri}, \citenamefont
  {Chancellor},\ and\ \citenamefont {Lidar}}]{q-sig}%
  \BibitemOpen
  \bibfield  {author} {\bibinfo {author} {\bibfnamefont {Sergio}\ \bibnamefont
  {Boixo}}, \bibinfo {author} {\bibfnamefont {Tameem}\ \bibnamefont {Albash}},
  \bibinfo {author} {\bibfnamefont {Federico~M.}\ \bibnamefont {Spedalieri}},
  \bibinfo {author} {\bibfnamefont {Nicholas}\ \bibnamefont {Chancellor}}, \
  and\ \bibinfo {author} {\bibfnamefont {Daniel~A.}\ \bibnamefont {Lidar}},\
  }\bibfield  {title} {\enquote {\bibinfo {title} {Experimental signature of
  programmable quantum annealing},}\ }\href {\doibase 10.1038/ncomms3067}
  {\bibfield  {journal} {\bibinfo  {journal} {Nat. Commun.}\ }\textbf {\bibinfo
  {volume} {4}},\ \bibinfo {pages} {2067} (\bibinfo {year} {2013})}\BibitemShut
  {NoStop}%
\bibitem [{\citenamefont {Albash}\ \emph {et~al.}(2015)\citenamefont {Albash},
  \citenamefont {Vinci}, \citenamefont {Mishra}, \citenamefont {Warburton},\
  and\ \citenamefont {Lidar}}]{q-sig2}%
  \BibitemOpen
  \bibfield  {author} {\bibinfo {author} {\bibfnamefont {Tameem}\ \bibnamefont
  {Albash}}, \bibinfo {author} {\bibfnamefont {Walter}\ \bibnamefont {Vinci}},
  \bibinfo {author} {\bibfnamefont {Anurag}\ \bibnamefont {Mishra}}, \bibinfo
  {author} {\bibfnamefont {Paul~A.}\ \bibnamefont {Warburton}}, \ and\ \bibinfo
  {author} {\bibfnamefont {Daniel~A.}\ \bibnamefont {Lidar}},\ }\bibfield
  {title} {\enquote {\bibinfo {title} {Consistency tests of classical and
  quantum models for a quantum annealer},}\ }\href
  {http://link.aps.org/doi/10.1103/PhysRevA.91.042314} {\bibfield  {journal}
  {\bibinfo  {journal} {Phys. Rev. A}\ }\textbf {\bibinfo {volume} {91}},\
  \bibinfo {pages} {042314--} (\bibinfo {year} {2015})}\BibitemShut {NoStop}%
\bibitem [{\citenamefont {Tolpygo}\ \emph
  {et~al.}(2015{\natexlab{a}})\citenamefont {Tolpygo}, \citenamefont
  {Bolkhovsky}, \citenamefont {Weir}, \citenamefont {Johnson}, \citenamefont
  {Gouker},\ and\ \citenamefont {Oliver}}]{LL1}%
  \BibitemOpen
  \bibfield  {author} {\bibinfo {author} {\bibfnamefont {S.~K.}\ \bibnamefont
  {Tolpygo}}, \bibinfo {author} {\bibfnamefont {V.}~\bibnamefont {Bolkhovsky}},
  \bibinfo {author} {\bibfnamefont {T.~J.}\ \bibnamefont {Weir}}, \bibinfo
  {author} {\bibfnamefont {L.~M.}\ \bibnamefont {Johnson}}, \bibinfo {author}
  {\bibfnamefont {M.~A.}\ \bibnamefont {Gouker}}, \ and\ \bibinfo {author}
  {\bibfnamefont {W.~D.}\ \bibnamefont {Oliver}},\ }\bibfield  {title}
  {\enquote {\bibinfo {title} {Fabrication process and properties of
  fully-planarized deep-submicron nb/al-alox/nb josephson junctions for vlsi
  circuits},}\ }\href {\doibase 10.1109/TASC.2014.2374836} {\bibfield
  {journal} {\bibinfo  {journal} {IEEE Transactions on Applied
  Superconductivity}\ }\textbf {\bibinfo {volume} {25}},\ \bibinfo {pages}
  {1--12} (\bibinfo {year} {2015}{\natexlab{a}})}\BibitemShut {NoStop}%
\bibitem [{\citenamefont {Tolpygo}\ \emph
  {et~al.}(2015{\natexlab{b}})\citenamefont {Tolpygo}, \citenamefont
  {Bolkhovsky}, \citenamefont {Weir}, \citenamefont {Galbraith}, \citenamefont
  {Johnson}, \citenamefont {Gouker},\ and\ \citenamefont {Semenov}}]{LL2}%
  \BibitemOpen
  \bibfield  {author} {\bibinfo {author} {\bibfnamefont {S.~K.}\ \bibnamefont
  {Tolpygo}}, \bibinfo {author} {\bibfnamefont {V.}~\bibnamefont {Bolkhovsky}},
  \bibinfo {author} {\bibfnamefont {T.~J.}\ \bibnamefont {Weir}}, \bibinfo
  {author} {\bibfnamefont {C.~J.}\ \bibnamefont {Galbraith}}, \bibinfo {author}
  {\bibfnamefont {L.~M.}\ \bibnamefont {Johnson}}, \bibinfo {author}
  {\bibfnamefont {M.~A.}\ \bibnamefont {Gouker}}, \ and\ \bibinfo {author}
  {\bibfnamefont {V.~K.}\ \bibnamefont {Semenov}},\ }\bibfield  {title}
  {\enquote {\bibinfo {title} {Inductance of circuit structures for mit ll
  superconductor electronics fabrication process with 8 niobium layers},}\
  }\href {\doibase 10.1109/TASC.2014.2369213} {\bibfield  {journal} {\bibinfo
  {journal} {IEEE Transactions on Applied Superconductivity}\ }\textbf
  {\bibinfo {volume} {25}},\ \bibinfo {pages} {1--5} (\bibinfo {year}
  {2015}{\natexlab{b}})}\BibitemShut {NoStop}%
\bibitem [{\citenamefont {Jin}\ \emph {et~al.}(2015)\citenamefont {Jin},
  \citenamefont {Kamal}, \citenamefont {Sears}, \citenamefont {Gudmundsen},
  \citenamefont {Hover}, \citenamefont {Miloshi}, \citenamefont {Slattery},
  \citenamefont {Yan}, \citenamefont {Yoder}, \citenamefont {Orlando},
  \citenamefont {Gustavsson},\ and\ \citenamefont {Oliver}}]{LL3}%
  \BibitemOpen
  \bibfield  {author} {\bibinfo {author} {\bibfnamefont {X.~Y.}\ \bibnamefont
  {Jin}}, \bibinfo {author} {\bibfnamefont {A.}~\bibnamefont {Kamal}}, \bibinfo
  {author} {\bibfnamefont {A.~P.}\ \bibnamefont {Sears}}, \bibinfo {author}
  {\bibfnamefont {T.}~\bibnamefont {Gudmundsen}}, \bibinfo {author}
  {\bibfnamefont {D.}~\bibnamefont {Hover}}, \bibinfo {author} {\bibfnamefont
  {J.}~\bibnamefont {Miloshi}}, \bibinfo {author} {\bibfnamefont
  {R.}~\bibnamefont {Slattery}}, \bibinfo {author} {\bibfnamefont
  {F.}~\bibnamefont {Yan}}, \bibinfo {author} {\bibfnamefont {J.}~\bibnamefont
  {Yoder}}, \bibinfo {author} {\bibfnamefont {T.~P.}\ \bibnamefont {Orlando}},
  \bibinfo {author} {\bibfnamefont {S.}~\bibnamefont {Gustavsson}}, \ and\
  \bibinfo {author} {\bibfnamefont {W.~D.}\ \bibnamefont {Oliver}},\ }\bibfield
   {title} {\enquote {\bibinfo {title} {Thermal and residual excited-state
  population in a 3d transmon qubit},}\ }\href {\doibase
  10.1103/PhysRevLett.114.240501} {\bibfield  {journal} {\bibinfo  {journal}
  {Phys. Rev. Lett.}\ }\textbf {\bibinfo {volume} {114}},\ \bibinfo {pages}
  {240501} (\bibinfo {year} {2015})}\BibitemShut {NoStop}%
\bibitem [{\citenamefont {Lanting}\ \emph {et~al.}(2017)\citenamefont
  {Lanting}, \citenamefont {King}, \citenamefont {Evert},\ and\ \citenamefont
  {Hoskinson}}]{PhysRevA.96.042322}%
  \BibitemOpen
  \bibfield  {author} {\bibinfo {author} {\bibfnamefont {Trevor}\ \bibnamefont
  {Lanting}}, \bibinfo {author} {\bibfnamefont {Andrew~D.}\ \bibnamefont
  {King}}, \bibinfo {author} {\bibfnamefont {Bram}\ \bibnamefont {Evert}}, \
  and\ \bibinfo {author} {\bibfnamefont {Emile}\ \bibnamefont {Hoskinson}},\
  }\bibfield  {title} {\enquote {\bibinfo {title} {Experimental demonstration
  of perturbative anticrossing mitigation using nonuniform driver
  hamiltonians},}\ }\href {\doibase 10.1103/PhysRevA.96.042322} {\bibfield
  {journal} {\bibinfo  {journal} {Phys. Rev. A}\ }\textbf {\bibinfo {volume}
  {96}},\ \bibinfo {pages} {042322} (\bibinfo {year} {2017})}\BibitemShut
  {NoStop}%
\bibitem [{\citenamefont {Perdomo-Ortiz}\ \emph {et~al.}(2011)\citenamefont
  {Perdomo-Ortiz}, \citenamefont {Venegas-Andraca},\ and\ \citenamefont
  {Aspuru-Guzik}}]{Perdomo-Ortiz:2011fh}%
  \BibitemOpen
  \bibfield  {author} {\bibinfo {author} {\bibfnamefont {Alejandro}\
  \bibnamefont {Perdomo-Ortiz}}, \bibinfo {author} {\bibfnamefont
  {Salvador~E.}\ \bibnamefont {Venegas-Andraca}}, \ and\ \bibinfo {author}
  {\bibfnamefont {Al{\'a}n}\ \bibnamefont {Aspuru-Guzik}},\ }\bibfield  {title}
  {\enquote {\bibinfo {title} {A study of heuristic guesses for adiabatic
  quantum computation},}\ }\href {\doibase 10.1007/s11128-010-0168-z}
  {\bibfield  {journal} {\bibinfo  {journal} {Quantum Information Processing}\
  }\textbf {\bibinfo {volume} {10}},\ \bibinfo {pages} {33--52} (\bibinfo
  {year} {2011})}\BibitemShut {NoStop}%
\bibitem [{\citenamefont {Roland}\ and\ \citenamefont
  {Cerf}(2002)}]{roland:02}%
  \BibitemOpen
  \bibfield  {author} {\bibinfo {author} {\bibfnamefont {J{\'e}r{\'e}mie}\
  \bibnamefont {Roland}}\ and\ \bibinfo {author} {\bibfnamefont {Nicolas~J.}\
  \bibnamefont {Cerf}},\ }\bibfield  {title} {\enquote {\bibinfo {title}
  {Quantum search by local adiabatic evolution},}\ }\href
  {http://link.aps.org/doi/10.1103/PhysRevA.65.042308} {\bibfield  {journal}
  {\bibinfo  {journal} {Phys. Rev. A}\ }\textbf {\bibinfo {volume} {65}},\
  \bibinfo {pages} {042308--} (\bibinfo {year} {2002})}\BibitemShut {NoStop}%
\bibitem [{\citenamefont {Hen}(2014{\natexlab{a}})}]{hen:14}%
  \BibitemOpen
  \bibfield  {author} {\bibinfo {author} {\bibfnamefont {Itay}\ \bibnamefont
  {Hen}},\ }\bibfield  {title} {\enquote {\bibinfo {title} {Continuous-time
  quantum algorithms for unstructured problems},}\ }\href {\doibase
  10.1088/1751-8113/47/4/045305} {\bibfield  {journal} {\bibinfo  {journal}
  {Journal of Physics A: Mathematical and Theoretical}\ }\textbf {\bibinfo
  {volume} {47}},\ \bibinfo {pages} {045305} (\bibinfo {year}
  {2014}{\natexlab{a}})}\BibitemShut {NoStop}%
\bibitem [{\citenamefont {Hen}(2014{\natexlab{b}})}]{hen:14b}%
  \BibitemOpen
  \bibfield  {author} {\bibinfo {author} {\bibfnamefont {Itay}\ \bibnamefont
  {Hen}},\ }\bibfield  {title} {\enquote {\bibinfo {title} {How fast can
  quantum annealers count?}}\ }\href
  {http://stacks.iop.org/1751-8121/47/i=23/a=235304} {\bibfield  {journal}
  {\bibinfo  {journal} {Journal of Physics A: Mathematical and Theoretical}\
  }\textbf {\bibinfo {volume} {47}},\ \bibinfo {pages} {235304} (\bibinfo
  {year} {2014}{\natexlab{b}})}\BibitemShut {NoStop}%
\bibitem [{\citenamefont {Hen}(2014{\natexlab{c}})}]{henPeriodFinding}%
  \BibitemOpen
  \bibfield  {author} {\bibinfo {author} {\bibfnamefont {I.}~\bibnamefont
  {Hen}},\ }\bibfield  {title} {\enquote {\bibinfo {title} {Period finding with
  adiabatic quantum computation},}\ }\href
  {http://stacks.iop.org/0295-5075/105/i=5/a=50005} {\bibfield  {journal}
  {\bibinfo  {journal} {EPL (Europhysics Letters)}\ }\textbf {\bibinfo {volume}
  {105}},\ \bibinfo {pages} {50005} (\bibinfo {year}
  {2014}{\natexlab{c}})}\BibitemShut {NoStop}%
\bibitem [{\citenamefont {Somma}\ \emph {et~al.}(2012)\citenamefont {Somma},
  \citenamefont {Nagaj},\ and\ \citenamefont {Kieferov\'a}}]{sommaGlued}%
  \BibitemOpen
  \bibfield  {author} {\bibinfo {author} {\bibfnamefont {Rolando~D.}\
  \bibnamefont {Somma}}, \bibinfo {author} {\bibfnamefont {Daniel}\
  \bibnamefont {Nagaj}}, \ and\ \bibinfo {author} {\bibfnamefont {M\'aria}\
  \bibnamefont {Kieferov\'a}},\ }\bibfield  {title} {\enquote {\bibinfo {title}
  {Quantum speedup by quantum annealing},}\ }\href {\doibase
  10.1103/PhysRevLett.109.050501} {\bibfield  {journal} {\bibinfo  {journal}
  {Phys. Rev. Lett.}\ }\textbf {\bibinfo {volume} {109}},\ \bibinfo {pages}
  {050501} (\bibinfo {year} {2012})}\BibitemShut {NoStop}%
\bibitem [{Note1()}]{Note1}%
  \BibitemOpen
  \bibinfo {note} {We exclude here algorithms derived via the polynomial
  equivalence theorem between the circuit and adiabatic models of quantum
  computing~\cite {aharonov_adiabatic_2007} since the `final' Hamiltonians are
  not necessarily diagonal in the computational basis.}\BibitemShut {Stop}%
\bibitem [{\citenamefont {{Farhi}}\ and\ \citenamefont
  {{Gutmann}}(1996)}]{analogAnalogue}%
  \BibitemOpen
  \bibfield  {author} {\bibinfo {author} {\bibfnamefont {E.}~\bibnamefont
  {{Farhi}}}\ and\ \bibinfo {author} {\bibfnamefont {S.}~\bibnamefont
  {{Gutmann}}},\ }\bibfield  {title} {\enquote {\bibinfo {title} {{An Analog
  Analogue of a Digital Quantum Computation}},}\ }\href@noop {} {\bibfield
  {journal} {\bibinfo  {journal} {eprint arXiv:quant-ph/9612026}\ } (\bibinfo
  {year} {1996})},\ \Eprint {http://arxiv.org/abs/quant-ph/9612026}
  {quant-ph/9612026} \BibitemShut {NoStop}%
\bibitem [{\citenamefont {Grover}(1997)}]{grover:97}%
  \BibitemOpen
  \bibfield  {author} {\bibinfo {author} {\bibfnamefont {Lov~K.}\ \bibnamefont
  {Grover}},\ }\bibfield  {title} {\enquote {\bibinfo {title} {Quantum
  mechanics helps in searching for a needle in a haystack},}\ }\href
  {http://link.aps.org/doi/10.1103/PhysRevLett.79.325} {\bibfield  {journal}
  {\bibinfo  {journal} {Phys. Rev. Lett.}\ }\textbf {\bibinfo {volume} {79}},\
  \bibinfo {pages} {325--328} (\bibinfo {year} {1997})}\BibitemShut {NoStop}%
\bibitem [{\citenamefont {Aaronson}(2005)}]{Aaronson}%
  \BibitemOpen
  \bibfield  {author} {\bibinfo {author} {\bibfnamefont {Scott}\ \bibnamefont
  {Aaronson}},\ }\bibfield  {title} {\enquote {\bibinfo {title} {Guest column:
  Np-complete problems and physical reality},}\ }\href {\doibase
  10.1145/1052796.1052804} {\bibfield  {journal} {\bibinfo  {journal} {SIGACT
  News}\ }\textbf {\bibinfo {volume} {36}},\ \bibinfo {pages} {30--52}
  (\bibinfo {year} {2005})}\BibitemShut {NoStop}%
\bibitem [{\citenamefont {Vergis}\ \emph {et~al.}(1986)\citenamefont {Vergis},
  \citenamefont {Steiglitz},\ and\ \citenamefont {Dickinson}}]{VERGIS198691}%
  \BibitemOpen
  \bibfield  {author} {\bibinfo {author} {\bibfnamefont {Anastasios}\
  \bibnamefont {Vergis}}, \bibinfo {author} {\bibfnamefont {Kenneth}\
  \bibnamefont {Steiglitz}}, \ and\ \bibinfo {author} {\bibfnamefont {Bradley}\
  \bibnamefont {Dickinson}},\ }\bibfield  {title} {\enquote {\bibinfo {title}
  {The complexity of analog computation},}\ }\href {\doibase
  https://doi.org/10.1016/0378-4754(86)90105-9} {\bibfield  {journal} {\bibinfo
   {journal} {Mathematics and Computers in Simulation}\ }\textbf {\bibinfo
  {volume} {28}},\ \bibinfo {pages} {91 -- 113} (\bibinfo {year}
  {1986})}\BibitemShut {NoStop}%
\bibitem [{\citenamefont {Jackson}(1960)}]{analogComputing}%
  \BibitemOpen
  \bibfield  {author} {\bibinfo {author} {\bibfnamefont {Albert~S.}\
  \bibnamefont {Jackson}},\ }\href@noop {} {\emph {\bibinfo {title} {Analog
  Computation}}}\ (\bibinfo  {publisher} {McGraw-Hill},\ \bibinfo {address}
  {New York, USA},\ \bibinfo {year} {1960})\BibitemShut {NoStop}%
\bibitem [{\citenamefont {Roland}\ and\ \citenamefont {Cerf}(2005)}]{RCNoise}%
  \BibitemOpen
  \bibfield  {author} {\bibinfo {author} {\bibfnamefont {J\'er\'emie}\
  \bibnamefont {Roland}}\ and\ \bibinfo {author} {\bibfnamefont {Nicolas~J.}\
  \bibnamefont {Cerf}},\ }\bibfield  {title} {\enquote {\bibinfo {title} {Noise
  resistance of adiabatic quantum computation using random matrix theory},}\
  }\href {\doibase 10.1103/PhysRevA.71.032330} {\bibfield  {journal} {\bibinfo
  {journal} {Phys. Rev. A}\ }\textbf {\bibinfo {volume} {71}},\ \bibinfo
  {pages} {032330} (\bibinfo {year} {2005})}\BibitemShut {NoStop}%
\bibitem [{\citenamefont {Mandr\`a}\ \emph {et~al.}(2015)\citenamefont
  {Mandr\`a}, \citenamefont {Guerreschi},\ and\ \citenamefont
  {Aspuru-Guzik}}]{PhysRevA.92.062320}%
  \BibitemOpen
  \bibfield  {author} {\bibinfo {author} {\bibfnamefont {Salvatore}\
  \bibnamefont {Mandr\`a}}, \bibinfo {author} {\bibfnamefont {Gian~Giacomo}\
  \bibnamefont {Guerreschi}}, \ and\ \bibinfo {author} {\bibfnamefont {Al\'an}\
  \bibnamefont {Aspuru-Guzik}},\ }\bibfield  {title} {\enquote {\bibinfo
  {title} {Adiabatic quantum optimization in the presence of discrete noise:
  Reducing the problem dimensionality},}\ }\href {\doibase
  10.1103/PhysRevA.92.062320} {\bibfield  {journal} {\bibinfo  {journal} {Phys.
  Rev. A}\ }\textbf {\bibinfo {volume} {92}},\ \bibinfo {pages} {062320}
  (\bibinfo {year} {2015})}\BibitemShut {NoStop}%
\bibitem [{\citenamefont {Wild}\ \emph {et~al.}(2016)\citenamefont {Wild},
  \citenamefont {Gopalakrishnan}, \citenamefont {Knap}, \citenamefont {Yao},\
  and\ \citenamefont {Lukin}}]{wild}%
  \BibitemOpen
  \bibfield  {author} {\bibinfo {author} {\bibfnamefont {Dominik~S.}\
  \bibnamefont {Wild}}, \bibinfo {author} {\bibfnamefont {Sarang}\ \bibnamefont
  {Gopalakrishnan}}, \bibinfo {author} {\bibfnamefont {Michael}\ \bibnamefont
  {Knap}}, \bibinfo {author} {\bibfnamefont {Norman~Y.}\ \bibnamefont {Yao}}, \
  and\ \bibinfo {author} {\bibfnamefont {Mikhail~D.}\ \bibnamefont {Lukin}},\
  }\bibfield  {title} {\enquote {\bibinfo {title} {Adiabatic quantum search in
  open systems},}\ }\href {\doibase 10.1103/PhysRevLett.117.150501} {\bibfield
  {journal} {\bibinfo  {journal} {Phys. Rev. Lett.}\ }\textbf {\bibinfo
  {volume} {117}},\ \bibinfo {pages} {150501} (\bibinfo {year}
  {2016})}\BibitemShut {NoStop}%
\bibitem [{\citenamefont {Amin}\ \emph {et~al.}(2009)\citenamefont {Amin},
  \citenamefont {Averin},\ and\ \citenamefont
  {Nesteroff}}]{PhysRevA.79.022107}%
  \BibitemOpen
  \bibfield  {author} {\bibinfo {author} {\bibfnamefont {M.~H.~S.}\
  \bibnamefont {Amin}}, \bibinfo {author} {\bibfnamefont {Dmitri~V.}\
  \bibnamefont {Averin}}, \ and\ \bibinfo {author} {\bibfnamefont {James~A.}\
  \bibnamefont {Nesteroff}},\ }\bibfield  {title} {\enquote {\bibinfo {title}
  {Decoherence in adiabatic quantum computation},}\ }\href {\doibase
  10.1103/PhysRevA.79.022107} {\bibfield  {journal} {\bibinfo  {journal} {Phys.
  Rev. A}\ }\textbf {\bibinfo {volume} {79}},\ \bibinfo {pages} {022107}
  (\bibinfo {year} {2009})}\BibitemShut {NoStop}%
\bibitem [{\citenamefont {Tiersch}\ and\ \citenamefont
  {Sch\"utzhold}(2007)}]{PhysRevA.75.062313}%
  \BibitemOpen
  \bibfield  {author} {\bibinfo {author} {\bibfnamefont {Markus}\ \bibnamefont
  {Tiersch}}\ and\ \bibinfo {author} {\bibfnamefont {Ralf}\ \bibnamefont
  {Sch\"utzhold}},\ }\bibfield  {title} {\enquote {\bibinfo {title}
  {Non-markovian decoherence in the adiabatic quantum search algorithm},}\
  }\href {\doibase 10.1103/PhysRevA.75.062313} {\bibfield  {journal} {\bibinfo
  {journal} {Phys. Rev. A}\ }\textbf {\bibinfo {volume} {75}},\ \bibinfo
  {pages} {062313} (\bibinfo {year} {2007})}\BibitemShut {NoStop}%
\bibitem [{\citenamefont {{\AA}berg}\ \emph
  {et~al.}(2005{\natexlab{a}})\citenamefont {{\AA}berg}, \citenamefont {Kult},\
  and\ \citenamefont {Sj{\"o}qvist}}]{PhysRevA.71.060312}%
  \BibitemOpen
  \bibfield  {author} {\bibinfo {author} {\bibfnamefont {Johan}\ \bibnamefont
  {{\AA}berg}}, \bibinfo {author} {\bibfnamefont {David}\ \bibnamefont {Kult}},
  \ and\ \bibinfo {author} {\bibfnamefont {Erik}\ \bibnamefont
  {Sj{\"o}qvist}},\ }\bibfield  {title} {\enquote {\bibinfo {title} {Robustness
  of the adiabatic quantum search},}\ }\href {\doibase
  10.1103/PhysRevA.71.060312} {\bibfield  {journal} {\bibinfo  {journal} {Phys.
  Rev. A}\ }\textbf {\bibinfo {volume} {71}},\ \bibinfo {pages} {060312}
  (\bibinfo {year} {2005}{\natexlab{a}})}\BibitemShut {NoStop}%
\bibitem [{\citenamefont {{\AA}berg}\ \emph
  {et~al.}(2005{\natexlab{b}})\citenamefont {{\AA}berg}, \citenamefont {Kult},\
  and\ \citenamefont {Sj{\"o}qvist}}]{PhysRevA.72.042317}%
  \BibitemOpen
  \bibfield  {author} {\bibinfo {author} {\bibfnamefont {Johan}\ \bibnamefont
  {{\AA}berg}}, \bibinfo {author} {\bibfnamefont {David}\ \bibnamefont {Kult}},
  \ and\ \bibinfo {author} {\bibfnamefont {Erik}\ \bibnamefont
  {Sj{\"o}qvist}},\ }\bibfield  {title} {\enquote {\bibinfo {title} {Quantum
  adiabatic search with decoherence in the instantaneous energy eigenbasis},}\
  }\href {\doibase 10.1103/PhysRevA.72.042317} {\bibfield  {journal} {\bibinfo
  {journal} {Phys. Rev. A}\ }\textbf {\bibinfo {volume} {72}},\ \bibinfo
  {pages} {042317} (\bibinfo {year} {2005}{\natexlab{b}})}\BibitemShut
  {NoStop}%
\bibitem [{\citenamefont {de~Vega}\ \emph {et~al.}(2010)\citenamefont
  {de~Vega}, \citenamefont {Ba{\~{n}}uls},\ and\ \citenamefont
  {P{\'{e}}rez}}]{de_Vega_2010}%
  \BibitemOpen
  \bibfield  {author} {\bibinfo {author} {\bibfnamefont {In{\'{e}}s}\
  \bibnamefont {de~Vega}}, \bibinfo {author} {\bibfnamefont {Mari~Carmen}\
  \bibnamefont {Ba{\~{n}}uls}}, \ and\ \bibinfo {author} {\bibfnamefont
  {A}~\bibnamefont {P{\'{e}}rez}},\ }\bibfield  {title} {\enquote {\bibinfo
  {title} {Effects of dissipation on an adiabatic quantum search algorithm},}\
  }\href {\doibase 10.1088/1367-2630/12/12/123010} {\bibfield  {journal}
  {\bibinfo  {journal} {New Journal of Physics}\ }\textbf {\bibinfo {volume}
  {12}},\ \bibinfo {pages} {123010} (\bibinfo {year} {2010})}\BibitemShut
  {NoStop}%
\bibitem [{\citenamefont {Mostame}\ \emph {et~al.}(2010)\citenamefont
  {Mostame}, \citenamefont {Schaller},\ and\ \citenamefont
  {Sch\"utzhold}}]{PhysRevA.81.032305}%
  \BibitemOpen
  \bibfield  {author} {\bibinfo {author} {\bibfnamefont {Sarah}\ \bibnamefont
  {Mostame}}, \bibinfo {author} {\bibfnamefont {Gernot}\ \bibnamefont
  {Schaller}}, \ and\ \bibinfo {author} {\bibfnamefont {Ralf}\ \bibnamefont
  {Sch\"utzhold}},\ }\bibfield  {title} {\enquote {\bibinfo {title}
  {Decoherence in a dynamical quantum phase transition},}\ }\href {\doibase
  10.1103/PhysRevA.81.032305} {\bibfield  {journal} {\bibinfo  {journal} {Phys.
  Rev. A}\ }\textbf {\bibinfo {volume} {81}},\ \bibinfo {pages} {032305}
  (\bibinfo {year} {2010})}\BibitemShut {NoStop}%
\bibitem [{\citenamefont {Aharonov}\ \emph {et~al.}(2006)\citenamefont
  {Aharonov}, \citenamefont {Kitaev},\ and\ \citenamefont
  {Preskill}}]{PhysRevLett.96.050504}%
  \BibitemOpen
  \bibfield  {author} {\bibinfo {author} {\bibfnamefont {Dorit}\ \bibnamefont
  {Aharonov}}, \bibinfo {author} {\bibfnamefont {Alexei}\ \bibnamefont
  {Kitaev}}, \ and\ \bibinfo {author} {\bibfnamefont {John}\ \bibnamefont
  {Preskill}},\ }\bibfield  {title} {\enquote {\bibinfo {title} {Fault-tolerant
  quantum computation with long-range correlated noise},}\ }\href {\doibase
  10.1103/PhysRevLett.96.050504} {\bibfield  {journal} {\bibinfo  {journal}
  {Phys. Rev. Lett.}\ }\textbf {\bibinfo {volume} {96}},\ \bibinfo {pages}
  {050504} (\bibinfo {year} {2006})}\BibitemShut {NoStop}%
\bibitem [{\citenamefont {Jansen}\ \emph {et~al.}(2007)\citenamefont {Jansen},
  \citenamefont {Ruskai},\ and\ \citenamefont {Seiler}}]{jansen:07}%
  \BibitemOpen
  \bibfield  {author} {\bibinfo {author} {\bibfnamefont {S.}~\bibnamefont
  {Jansen}}, \bibinfo {author} {\bibfnamefont {M.~B.}\ \bibnamefont {Ruskai}},
  \ and\ \bibinfo {author} {\bibfnamefont {R.}~\bibnamefont {Seiler}},\
  }\bibfield  {title} {\enquote {\bibinfo {title} {Bounds for the adiabatic
  approximation with applications to quantum computation},}\ }\href {\doibase
  10.1063/1.2798382} {\bibfield  {journal} {\bibinfo  {journal} {J. Math.
  Phys.}\ }\textbf {\bibinfo {volume} {48}},\ \bibinfo {pages} {102111}
  (\bibinfo {year} {2007})}\BibitemShut {NoStop}%
\bibitem [{\citenamefont {Lidar}\ \emph {et~al.}(2009)\citenamefont {Lidar},
  \citenamefont {Rezakhani},\ and\ \citenamefont {Hamma}}]{lidarGap}%
  \BibitemOpen
  \bibfield  {author} {\bibinfo {author} {\bibfnamefont {D.~A.}\ \bibnamefont
  {Lidar}}, \bibinfo {author} {\bibfnamefont {A.~T.}\ \bibnamefont
  {Rezakhani}}, \ and\ \bibinfo {author} {\bibfnamefont {A.}~\bibnamefont
  {Hamma}},\ }\bibfield  {title} {\enquote {\bibinfo {title} {Adiabatic
  approximation with exponential accuracy for many-body systems and quantum
  computation},}\ }\href {\doibase http://dx.doi.org/10.1063/1.3236685}
  {\bibfield  {journal} {\bibinfo  {journal} {J. Math. Phys.}\ }\textbf
  {\bibinfo {volume} {50}},\ \bibinfo {eid} {102106} (\bibinfo {year}
  {2009})}\BibitemShut {NoStop}%
\bibitem [{\citenamefont {Kato}(1950)}]{kato:51}%
  \BibitemOpen
  \bibfield  {author} {\bibinfo {author} {\bibfnamefont {T.}~\bibnamefont
  {Kato}},\ }\bibfield  {title} {\enquote {\bibinfo {title} {On the adiabatic
  theorem of quantum mechanics},}\ }\href {\doibase 10.1143/JPSJ.5.435}
  {\bibfield  {journal} {\bibinfo  {journal} {J. Phys. Soc. Jap.}\ }\textbf
  {\bibinfo {volume} {5}},\ \bibinfo {pages} {435--439} (\bibinfo {year}
  {1950})}\BibitemShut {NoStop}%
\bibitem [{\citenamefont {Farhi}\ \emph {et~al.}(2008)\citenamefont {Farhi},
  \citenamefont {Goldstone}, \citenamefont {Gutmann},\ and\ \citenamefont
  {Nagaj}}]{farhi:08}%
  \BibitemOpen
  \bibfield  {author} {\bibinfo {author} {\bibfnamefont {E.}~\bibnamefont
  {Farhi}}, \bibinfo {author} {\bibfnamefont {J.}~\bibnamefont {Goldstone}},
  \bibinfo {author} {\bibfnamefont {S.}~\bibnamefont {Gutmann}}, \ and\
  \bibinfo {author} {\bibfnamefont {D}~\bibnamefont {Nagaj}},\ }\bibfield
  {title} {\enquote {\bibinfo {title} {How to make the quantum adiabatic
  algorithm fail},}\ }\href@noop {} {\bibfield  {journal} {\bibinfo  {journal}
  {International Journal of Quantum Information}\ }\textbf {\bibinfo {volume}
  {6}},\ \bibinfo {pages} {503} (\bibinfo {year} {2008})},\ \bibinfo {note}
  {(arXiv:quant-ph/0512159)}\BibitemShut {NoStop}%
\bibitem [{\citenamefont {Roland}\ and\ \citenamefont
  {Cerf}(2003)}]{roland:03}%
  \BibitemOpen
  \bibfield  {author} {\bibinfo {author} {\bibfnamefont {J\'er\'emie}\
  \bibnamefont {Roland}}\ and\ \bibinfo {author} {\bibfnamefont {Nicolas~J.}\
  \bibnamefont {Cerf}},\ }\bibfield  {title} {\enquote {\bibinfo {title}
  {Adiabatic quantum search algorithm for structured problems},}\ }\href
  {\doibase 10.1103/PhysRevA.68.062312} {\bibfield  {journal} {\bibinfo
  {journal} {Phys. Rev. A}\ }\textbf {\bibinfo {volume} {68}},\ \bibinfo
  {pages} {062312} (\bibinfo {year} {2003})}\BibitemShut {NoStop}%
\bibitem [{\citenamefont {Childs}\ and\ \citenamefont
  {Goldstone}(2004)}]{childsGoldstone}%
  \BibitemOpen
  \bibfield  {author} {\bibinfo {author} {\bibfnamefont {Andrew~M.}\
  \bibnamefont {Childs}}\ and\ \bibinfo {author} {\bibfnamefont {Jeffrey}\
  \bibnamefont {Goldstone}},\ }\bibfield  {title} {\enquote {\bibinfo {title}
  {Spatial search by quantum walk},}\ }\href {\doibase
  10.1103/PhysRevA.70.022314} {\bibfield  {journal} {\bibinfo  {journal} {Phys.
  Rev. A}\ }\textbf {\bibinfo {volume} {70}},\ \bibinfo {pages} {022314}
  (\bibinfo {year} {2004})}\BibitemShut {NoStop}%
\bibitem [{\citenamefont {Hen}(2017)}]{realizableAQCsearch}%
  \BibitemOpen
  \bibfield  {author} {\bibinfo {author} {\bibfnamefont {I.}~\bibnamefont
  {Hen}},\ }\bibfield  {title} {\enquote {\bibinfo {title} {Realizable quantum
  adiabatic search},}\ }\href {http://stacks.iop.org/0295-5075/118/i=3/a=30003}
  {\bibfield  {journal} {\bibinfo  {journal} {EPL (Europhysics Letters)}\
  }\textbf {\bibinfo {volume} {118}},\ \bibinfo {pages} {30003} (\bibinfo
  {year} {2017})}\BibitemShut {NoStop}%
\bibitem [{\citenamefont {Press}\ \emph {et~al.}(1992)\citenamefont {Press},
  \citenamefont {Teukolsky}, \citenamefont {Vetterling},\ and\ \citenamefont
  {Flannery}}]{press:92}%
  \BibitemOpen
  \bibfield  {author} {\bibinfo {author} {\bibfnamefont {W.~H.}\ \bibnamefont
  {Press}}, \bibinfo {author} {\bibfnamefont {S.~A.}\ \bibnamefont
  {Teukolsky}}, \bibinfo {author} {\bibfnamefont {W.~T.}\ \bibnamefont
  {Vetterling}}, \ and\ \bibinfo {author} {\bibfnamefont {B.~P.}\ \bibnamefont
  {Flannery}},\ }\href@noop {} {\emph {\bibinfo {title} {Numerical Recipes in
  C, 2nd Ed.}}}\ (\bibinfo  {publisher} {Cambridge University Press},\ \bibinfo
  {address} {Cambridge},\ \bibinfo {year} {1992})\BibitemShut {NoStop}%
\bibitem [{\citenamefont {Cash}\ and\ \citenamefont {Karp}(1990)}]{cash:90}%
  \BibitemOpen
  \bibfield  {author} {\bibinfo {author} {\bibfnamefont {Jeff~R}\ \bibnamefont
  {Cash}}\ and\ \bibinfo {author} {\bibfnamefont {Alan~H}\ \bibnamefont
  {Karp}},\ }\bibfield  {title} {\enquote {\bibinfo {title} {A variable order
  runge-kutta method for initial value problems with rapidly varying right-hand
  sides},}\ }\href {\doibase 10.1145/79505.79507} {\bibfield  {journal}
  {\bibinfo  {journal} {ACM Trans. Math. Softw.}\ }\textbf {\bibinfo {volume}
  {16}},\ \bibinfo {pages} {201--222} (\bibinfo {year} {1990})}\BibitemShut
  {NoStop}%
\bibitem [{\citenamefont {Albash}\ and\ \citenamefont
  {Lidar}(2015)}]{PhysRevA.91.062320}%
  \BibitemOpen
  \bibfield  {author} {\bibinfo {author} {\bibfnamefont {Tameem}\ \bibnamefont
  {Albash}}\ and\ \bibinfo {author} {\bibfnamefont {Daniel~A.}\ \bibnamefont
  {Lidar}},\ }\bibfield  {title} {\enquote {\bibinfo {title} {Decoherence in
  adiabatic quantum computation},}\ }\href {\doibase
  10.1103/PhysRevA.91.062320} {\bibfield  {journal} {\bibinfo  {journal} {Phys.
  Rev. A}\ }\textbf {\bibinfo {volume} {91}},\ \bibinfo {pages} {062320}
  (\bibinfo {year} {2015})}\BibitemShut {NoStop}%
\bibitem [{\citenamefont {Albash}\ \emph {et~al.}(2012)\citenamefont {Albash},
  \citenamefont {Boixo}, \citenamefont {Lidar},\ and\ \citenamefont
  {Zanardi}}]{Albash_2012}%
  \BibitemOpen
  \bibfield  {author} {\bibinfo {author} {\bibfnamefont {Tameem}\ \bibnamefont
  {Albash}}, \bibinfo {author} {\bibfnamefont {Sergio}\ \bibnamefont {Boixo}},
  \bibinfo {author} {\bibfnamefont {Daniel~A}\ \bibnamefont {Lidar}}, \ and\
  \bibinfo {author} {\bibfnamefont {Paolo}\ \bibnamefont {Zanardi}},\
  }\bibfield  {title} {\enquote {\bibinfo {title} {Quantum adiabatic markovian
  master equations},}\ }\href {\doibase 10.1088/1367-2630/14/12/123016}
  {\bibfield  {journal} {\bibinfo  {journal} {New Journal of Physics}\ }\textbf
  {\bibinfo {volume} {14}},\ \bibinfo {pages} {123016} (\bibinfo {year}
  {2012})}\BibitemShut {NoStop}%
\bibitem [{\citenamefont {{Jarret}}\ \emph {et~al.}(2018)\citenamefont
  {{Jarret}}, \citenamefont {{Lackey}}, \citenamefont {{Liu}},\ and\
  \citenamefont {{Wan}}}]{Jarret2018}%
  \BibitemOpen
  \bibfield  {author} {\bibinfo {author} {\bibfnamefont {Michael}\ \bibnamefont
  {{Jarret}}}, \bibinfo {author} {\bibfnamefont {Brad}\ \bibnamefont
  {{Lackey}}}, \bibinfo {author} {\bibfnamefont {Aike}\ \bibnamefont {{Liu}}},
  \ and\ \bibinfo {author} {\bibfnamefont {Kianna}\ \bibnamefont {{Wan}}},\
  }\bibfield  {title} {\enquote {\bibinfo {title} {{Quantum adiabatic
  optimization without heuristics}},}\ }\href@noop {} {\bibfield  {journal}
  {\bibinfo  {journal} {arXiv e-prints}\ ,\ \bibinfo {eid} {arXiv:1810.04686}}
  (\bibinfo {year} {2018})},\ \Eprint {http://arxiv.org/abs/1810.04686}
  {arXiv:1810.04686 [quant-ph]} \BibitemShut {NoStop}%
\bibitem [{Note2()}]{Note2}%
  \BibitemOpen
  \bibinfo {note} {We thank Michael Jarret for pointing this out.}\BibitemShut
  {Stop}%
\bibitem [{\citenamefont {Cleve}\ \emph {et~al.}(2009)\citenamefont {Cleve},
  \citenamefont {Gottesman}, \citenamefont {Mosca}, \citenamefont {Somma},\
  and\ \citenamefont {Yonge-Mallo}}]{Cleve:2009:EDS:1536414.1536471}%
  \BibitemOpen
  \bibfield  {author} {\bibinfo {author} {\bibfnamefont {Richard}\ \bibnamefont
  {Cleve}}, \bibinfo {author} {\bibfnamefont {Daniel}\ \bibnamefont
  {Gottesman}}, \bibinfo {author} {\bibfnamefont {Michele}\ \bibnamefont
  {Mosca}}, \bibinfo {author} {\bibfnamefont {Rolando~D.}\ \bibnamefont
  {Somma}}, \ and\ \bibinfo {author} {\bibfnamefont {David}\ \bibnamefont
  {Yonge-Mallo}},\ }\bibfield  {title} {\enquote {\bibinfo {title} {Efficient
  discrete-time simulations of continuous-time quantum query algorithms},}\
  }in\ \href {\doibase 10.1145/1536414.1536471} {\emph {\bibinfo {booktitle}
  {Proceedings of the Forty-first Annual ACM Symposium on Theory of
  Computing}}},\ \bibinfo {series and number} {STOC '09}\ (\bibinfo
  {publisher} {ACM},\ \bibinfo {address} {New York, NY, USA},\ \bibinfo {year}
  {2009})\ pp.\ \bibinfo {pages} {409--416}\BibitemShut {NoStop}%
\bibitem [{\citenamefont {Aharonov}\ \emph {et~al.}(2007)\citenamefont
  {Aharonov}, \citenamefont {van Dam}, \citenamefont {Kempe}, \citenamefont
  {Landau}, \citenamefont {Lloyd},\ and\ \citenamefont
  {Regev}}]{aharonov_adiabatic_2007}%
  \BibitemOpen
  \bibfield  {author} {\bibinfo {author} {\bibfnamefont {Dorit}\ \bibnamefont
  {Aharonov}}, \bibinfo {author} {\bibfnamefont {Wim}\ \bibnamefont {van Dam}},
  \bibinfo {author} {\bibfnamefont {Julia}\ \bibnamefont {Kempe}}, \bibinfo
  {author} {\bibfnamefont {Zeph}\ \bibnamefont {Landau}}, \bibinfo {author}
  {\bibfnamefont {Seth}\ \bibnamefont {Lloyd}}, \ and\ \bibinfo {author}
  {\bibfnamefont {Oded}\ \bibnamefont {Regev}},\ }\bibfield  {title} {\enquote
  {\bibinfo {title} {Adiabatic quantum computation is equivalent to standard
  quantum computation},}\ }\href
  {http://epubs.siam.org/doi/abs/10.1137/S0097539705447323} {\bibfield
  {journal} {\bibinfo  {journal} {SIAM J. Comput.}\ }\textbf {\bibinfo {volume}
  {37}},\ \bibinfo {pages} {166--194} (\bibinfo {year} {2007})}\BibitemShut
  {NoStop}%
\end{thebibliography}
%

\end{document}